\pdfoutput=1
\documentclass[%
 reprint,
 superscriptaddress,
nofootinbib,
 amsmath,amssymb,
 aps,
 prd,
 floatfix
]{revtex4-1}

\usepackage{graphicx}
\usepackage{dcolumn}
\usepackage{bm}
\usepackage{amssymb}
\usepackage{color}
\usepackage[caption=false]{subfig}

\begin{document}

\preprint{APS/123-QED}

\title{Identification and mitigation of narrow spectral artifacts that degrade searches for persistent gravitational waves in the first two observing runs of Advanced LIGO}

\author{%
P.~B.~Covas,$^{1}$  
A.~Effler,$^{2}$  
E.~Goetz,$^{3,4}$  
P.~M.~Meyers,$^{5}$  
A.~Neunzert,$^{3}$  
M.~Oliver,$^{1}$  
B.~L.~Pearlstone,$^{6}$  
V.~J.~Roma,$^{7}$  
R.~M.~S.~Schofield,$^{7}$  
V.~B.~Adya,$^{4}$
% ...
\\
\medskip
(LSC Instrument Authors) 
\\
\medskip
}\noaffiliation

\affiliation {Universitat de les Illes Balears, IAC3---IEEC, E-07122 Palma de Mallorca, Spain }
\affiliation {LIGO Livingston Observatory, Livingston, LA 70754, USA }
\affiliation {University of Michigan, Ann Arbor, MI 48109, USA }
\affiliation {Max Planck Institute for Gravitational Physics (Albert Einstein Institute), D-30167 Hannover, Germany }
\affiliation {University of Minnesota, Minneapolis, MN 55455, USA }
\affiliation {SUPA, University of Glasgow, Glasgow G12 8QQ, United Kingdom }
\affiliation {University of Oregon, Eugene, OR 97403, USA }
\affiliation {INFN, Sezione di Roma, I-00185 Roma, Italy }
\affiliation {The Pennsylvania State University, University Park, PA 16802, USA }
\affiliation {OzGrav, School of Physics \& Astronomy, Monash University, Clayton 3800, Victoria, Australia }
\affiliation {LIGO, California Institute of Technology, Pasadena, CA 91125, USA }
\affiliation {Carleton College, Northfield, MN 55057, USA }
\affiliation {Artemis, Universit\'e C\^ote d'Azur, Observatoire C\^ote d'Azur, CNRS, CS 34229, F-06304 Nice Cedex 4, France }
\affiliation {Universit\`a di Roma `La Sapienza,' I-00185 Roma, Italy }
\affiliation {Bellevue College, Bellevue, WA 98007, USA }
\affiliation {LIGO Hanford Observatory, Richland, WA 99352, USA }
\affiliation {Louisiana State University, Baton Rouge, LA 70803, USA }
\affiliation {Syracuse University, Syracuse, NY 13244, USA }
\affiliation {LIGO, Massachusetts Institute of Technology, Cambridge, MA 02139, USA }
\affiliation {University of Florida, Gainesville, FL 32611, USA }
\affiliation {Nicolaus Copernicus Astronomical Center, Polish Academy of Sciences, 00-716, Warsaw, Poland }
\affiliation {OzGrav, University of Adelaide, Adelaide, South Australia 5005, Australia }
\affiliation {OzGrav, University of Melbourne, Parkville, Victoria 3010, Australia }
\affiliation {University of Birmingham, Birmingham B15 2TT, United Kingdom }
\affiliation {Columbia University, New York, NY 10027, USA }
\affiliation {Leibniz Universit\"at Hannover, D-30167 Hannover, Germany }
\affiliation {The University of Sheffield, Sheffield S10 2TN, United Kingdom }
\affiliation {Stanford University, Stanford, CA 94305, USA }
\affiliation {California State University, Los Angeles, 5151 State University Dr, Los Angeles, CA 90032, USA }
\affiliation {The University of Mississippi, University, MS 38677, USA }
\affiliation {Instituto Nacional de Pesquisas Espaciais, 12227-010 S\~{a}o Jos\'{e} dos Campos, S\~{a}o Paulo, Brazil }
\affiliation {Canadian Institute for Theoretical Astrophysics, University of Toronto, Toronto, Ontario M5S 3H8, Canada }
\affiliation {American University, Washington, D.C. 20016, USA }
\affiliation {Texas Tech University, Lubbock, TX 79409, USA }
\affiliation {OzGrav, Australian National University, Canberra, Australian Capital Territory 0200, Australia }
\affiliation {OzGrav, University of Western Australia, Crawley, Western Australia 6009, Australia }
\affiliation {SUPA, University of Strathclyde, Glasgow G1 1XQ, United Kingdom }
\affiliation {Cardiff University, Cardiff CF24 3AA, United Kingdom }
\affiliation {University of Washington Bothell, 18115 Campus Way NE, Bothell, WA 98011, USA }
\affiliation {Inter-University Centre for Astronomy and Astrophysics, Pune 411007, India }
\affiliation {Hobart and William Smith Colleges, Geneva, NY 14456, USA }
\affiliation {University of Sannio at Benevento, I-82100 Benevento, Italy and INFN, Sezione di Napoli, I-80100 Napoli, Italy }
\affiliation {Faculty of Physics, Lomonosov Moscow State University, Moscow 119991, Russia }
\affiliation {The University of Texas Rio Grande Valley, Brownsville, TX 78520, USA }
\affiliation {Institute of Physics, E\"otv\"os University, P\'azm\'any P. s. 1/A, Budapest 1117, Hungary }
\affiliation {SUPA, University of the West of Scotland, Paisley PA1 2BE, United Kingdom }
\affiliation {University of Washington, Seattle, WA 98195, USA }
\affiliation {California State University Fullerton, Fullerton, CA 92831, USA }
\affiliation {Kenyon College, Gambier, OH 43022, USA }
\affiliation {Rochester Institute of Technology, Rochester, NY 14623, USA }
\affiliation {University of Maryland, College Park, MD 20742, USA }


\begin{abstract}
Searches are under way in Advanced LIGO and Virgo data for persistent gravitational waves from continuous sources, e.g. rapidly rotating galactic neutron stars, and stochastic sources, e.g. relic gravitational waves from the Big Bang or superposition of distant astrophysical events such as mergers of black holes or neutron stars. These searches can be degraded by the presence of narrow spectral artifacts (lines) due to instrumental or environmental disturbances.
We describe a variety of methods used for finding, identifying and mitigating these artifacts, illustrated with particular examples. Results are provided in the form of lists of line artifacts that can safely be treated as non-astrophysical. Such lists are used to improve the efficiencies and sensitivities of continuous and stochastic gravitational wave searches by allowing vetoes of false outliers and permitting data cleaning. 
  
\end{abstract}

\maketitle

\section{\label{sec:intr}Introduction}

The recent detections of transient gravitational waves (GWs) from the merger of binary black holes and of binary neutron stars opened a new field of observational GW astronomy~\cite{FirstDetection,NSDetection}. 
The near future may also bring the discovery by the LIGO and Virgo detectors of {\it persistent} gravitational waves.
 
Persistent sources of long-duration GWs can be broadly classified as continuous wave (CW) sources, which have a deterministic phase evolution, and a stochastic gravitational-wave background (SGWB), for which the signal is intrinsically random. The canonical sources for CWs (see~\cite{CWReview} for a review) are non-axisymmetric rotating neutron stars, emitting long-lasting and nearly monochromatic waves. When observed from Earth, these waves will be frequency-modulated due to the Doppler effect produced by the daily rotation and orbital motion of the Earth around the Sun. The SGWB is a superposition of many astrophysical and cosmological GW sources. Astrophysical sources are reviewed in~\cite{Regimbau_2011}. Cosmological sources of the SGWB include cosmic string networks~\cite{2005PhRvD..71f3510D,1976JPhA....9.1387K,2002PhLB..536..185S,2007PhRvL..98k1101S}, inflation~\cite{1994PhRvD..50.1157B,1979JETPL..30..682S,2007PhRvL..99v1301E,2012PhRvD..85b3525B,2012PhRvD..85b3534C,2013arXiv1305.5855L,1997PhRvD..55..435T,2006JCAP...04..010E}, phase transitions~\cite{Kosowsky1992rz,Kamionkowski:1993fg,thumb}, and the pre-Big Bang scenario~\cite{Gasperini:1992em,Gasperini:1993hu,PBBpaper,Gasperini:2016gre}. For reviews of search methods for the SGWB, see~\cite{RomanoCornish,Allen_Romano_1999}.

CW and SGWB searches look for long-duration signals, and are affected by different types of noise than those affecting short-duration searches. While compact binary coalescence and burst searches are degraded mainly by short-duration glitches (such as those described in~\cite{ALIGOTransient,ALIGODQO1Transient,GW150914TransientNoise}), CW and SGWB searches are mainly affected by long-lived peaks in the frequency spectra, especially narrow peaks, typically referred to as lines. CW searches can be degraded because their signals are intrinsically highly narrow-band, while SGWB searches can be degraded because of the tendency of a subset of instrumental lines in the two detectors to lie so close to each other that they exhibit spurious coherence between the detectors.

This problem presents two main detector characterization tasks for long-duration searches: first, to \emph{identify} line artifacts that are non-astrophysical in origin, allowing them to be flagged as noise; and, second, to determine the \emph{cause} of those artifacts when possible in order to guide efforts to remove them at the detector sites.
Spectral lines that affect the CW and the SGWB searches are typically quite narrow (high Q-factor, i.e., the ratio of peak frequency to line width) during a given coherent integration time. This focuses investigations for noise sources onto electronic components and mechanical components with high Q-factor resonances, and eliminating, for example, mechanical components with damped mechanical resonances.

In this report, we describe tools and methods used for data quality investigations relevant to long-duration searches, and provide examples of issues faced in the first two Advanced LIGO observing runs, O1 and O2. The paper is organized as follows: section~\ref{sec:discussion} summarizes the effects that noise has on the searches for persistent GWs; section~\ref{sec:data} briefly introduces LIGO data and noise sources; section~\ref{sec:coupling} gives examples of different noise coupling mechanisms to the GW channel; section~\ref{sec:tools} summarizes data analysis tools used for noise characterization; section~\ref{sec:examples} presents results from noise sources that were investigated and mitigated during O1 and O2; and section VII describes the procedures used to generate line lists for vetoing noise outliers.

Finally, we note that all of the methods presented here can be applied to both LIGO and Virgo detectors. We will focus, however, on data quality applied to the LIGO detectors only, as, at the time of this writing, there is significantly more Advanced LIGO observational data, which is needed for persistent GW searches.

\section{\label{sec:discussion}Effects of noise on CW and SGWB searches}
Spectral artifacts can degrade analyses that search for long-duration signals in different ways. Artifacts can lead search pipelines to return spurious outliers, which require laborious follow-up. Furthermore, if there is a putative GW signal at a frequency corresponding or nearby to a spectral artifact, then the signal power is obscured. For those analyses that rely on combining data from different detectors (e.g. cross-correlation or coherently combining data), then detection of signals overlapping with common detector artifacts may be impossible. On the other hand, some searches may be able to cope with an artifact if it occurs in just one detector.

Continuous GWs from spinning neutron stars are nearly monochromatic, with nearly constant signal frequency in the Solar System barycenter. When projected into the frame of a detector located on Earth, the signal is Doppler shifted into many frequency bins. Conversely, a narrow, stationary spectral artifact in the detector frame will impact many frequency bins when data is projected into the frame of the Solar System barycenter. For searches of a signal from a known pulsar with a given ephemeris, the impact of these artifacts is less than the impact on an all-sky search for unknown neutron stars (which may also be located in a binary system). In extreme cases, an all-sky search may be blind to a wide region of parameter space for a particular frequency range.

Searches for a stochastic GW background rely on cross-correlating GW strain channel data from multiple detectors and looking for excess power. Excess cross-correlation requires a stable phase between the two channels at a given frequency, and, thus, many single-detector artifacts are not found in the cross-correlation analysis. Correlated noise that causes excess power in the cross-correlation analysis, however, is excised from the analysis entirely by setting that frequency bin to zero before integration in the case of the standard search for a broadband SGWB. This reduces the search  sensitivity by a factor $\sim \sqrt{\frac{N_{b}}{N_{a}}}$ where $N_b$ is the number of frequency bins before notching and $N_a$ is the number of frequency bins after notching. In directed, narrow-band searches~\cite{stoch_directional_O1} we do not search for GWs at frequencies of known instrumental lines.

For both CW and SGWB searches, lists of known instrumental artifacts are created following the end of an observing run (further details are provided in section~\ref{sec:knownlines}). Then, depending on the search, these lists are used to: 1) clean the data before analysis by removing the affected data in the Fourier domain and replacing it with Gaussian noise measured in the nearby frequency bins; 2) avoid specific frequencies in analyses that are impacted by the artifacts; or 3) reject outliers that are clearly caused by the detector artifacts. This lets the analysis focus computational resources on regions of parameter space that are not degraded by spectral features. If a search pipeline returns a signal candidate which does not coincide with any known artifact, more detailed investigations are needed in order to assert that the signal cannot be produced by an artifact. 

\section{\label{sec:data}LIGO data and noise sources for searches of persistent gravitational waves}
The first Advanced LIGO observing run (O1) took place between 12 September 2015 and 12 January 2016, while the second Advanced LIGO observing run (O2) took place between 30 November 2016 and 25 August 2017. The Advanced LIGO detectors are located in Hanford, Washington (H1), and Livingston, Louisiana (L1).
The LIGO detectors are dual-recycled Michelson interferometers with Fabry-Perot arm cavities of $\approx$4 km (see \cite{aligo} for a review of the Advanced LIGO detectors configuration).

\begin{figure}[tbp]
\includegraphics[width=\columnwidth]{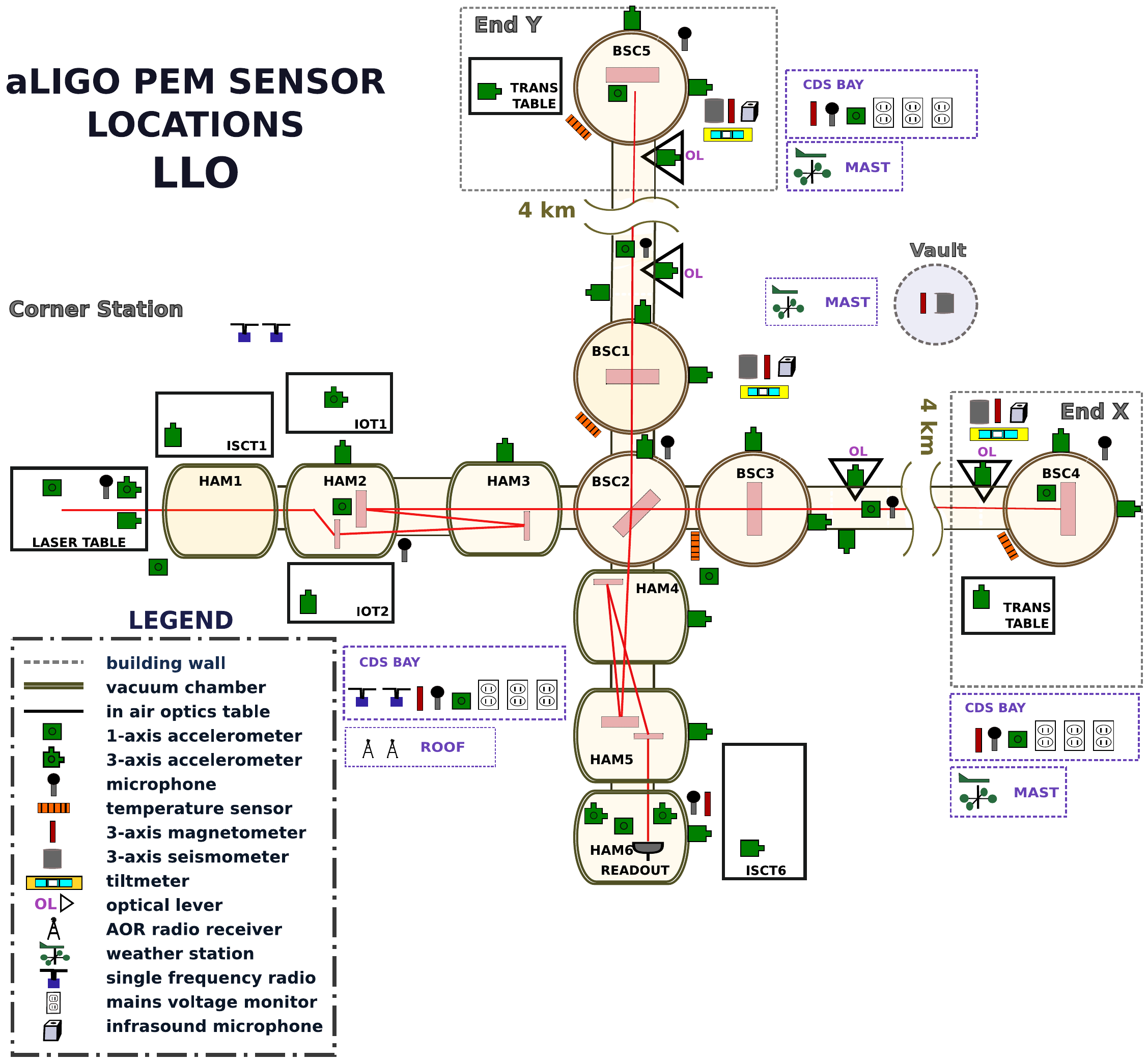}
\caption{\label{fig:PEMLLO}Locations of most auxiliary sensors at LIGO Livingston Observatory (LHO shares a similar layout). The gray dashed lines separate the End X and End Y Stations, which are located at the end of the 4~km arms, from the Corner Station building, located at the vertex of the detector. All stations contain an electronics room (encased by purple points in the diagram), where the computers that control the interferometer are housed.}
\end{figure}

LIGO detector data is typically characterized as stationary, Gaussian noise, but non-Gaussian detector artifacts are also present in LIGO data, e.g., occasional, short-duration transients (``glitches'') and long-duration narrow lines. Searches for transient GW signals will usually avoid analyzing times when a glitch occurs, while searches for persistent GW signals avoid analyzing data in frequency bands where narrow lines are present. This enables either type of search to consider the detector noise data to be essentially Gaussian.

While most lines in detector data are stationary, some of the lines have time-varying behavior (often called wandering lines), which can degrade detector sensitivity over a larger range of frequencies and increase the difficulty of distinguishing these artifacts from astrophysical signals when searching for a persistent signal from different sky locations. Some lines occur in a distinct pattern known as a \emph{comb}, with even-spacing in frequency between each tooth (each single line) of the comb. Tooth frequencies are given by $f_n = f_{o} + n*\delta f$, where $f_{o}$ is the offset (from 0~Hz) of the comb, $\delta f$ is the spacing, and $n$ is an integer. These combs are associated with linear or non-linear coupling of non-sinusoidal sources or with non-linear coupling of sinusoidal sources. A comb can also be recognized by a common time-dependent behavior of the teeth in the comb. The Fourier coefficients of a comb in the frequency domain can be used to uncover the time-domain waveform and help identify the source of the comb.

Lines and combs can have time-dependent behavior as the configuration of the detector changes, especially during periods of commissioning and maintenance. Some lines and combs have high amplitude and can be identified using only a short amount of data. Others have low amplitude and may only become evident with long integration time. Long integration time is also useful to better constrain the central frequency and width of a given line or to find the spacing of a comb.

A schematic diagram showing locations of vacuum chambers, main interferometer optics, and most of the Physical Environment Monitoring (PEM) sensors of the L1 detector can be seen in figure~\ref{fig:PEMLLO} (H1 has a similar layout). PEM sensors include, for example, accelerometers, microphones, temperature sensors, magnetometers, seismometers, etc. PEM sensors, particularly magnetometers, are often helpful in determining the causes of narrow spectral artifacts because they witness local noise sources that may couple to the main GW channel, and the PEM sensors do not witness GW signals (except in cases of complicated cross-coupling mechanisms, which can be identified using signal injections). Other auxiliary channels may also be useful in the same way.

\begin{figure}[tbp]
\includegraphics[width=1.0\columnwidth]{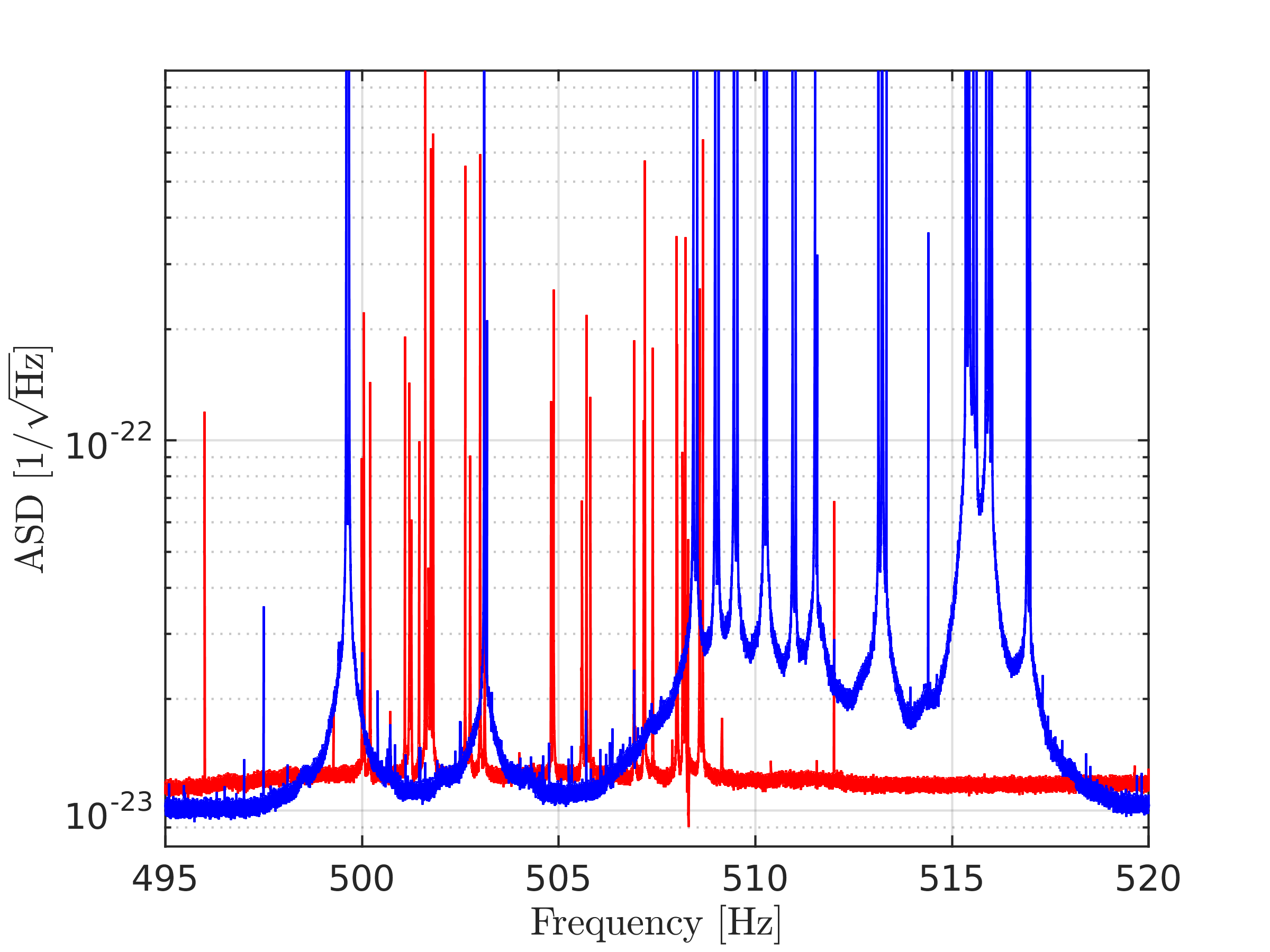}
\caption{\label{fig:ViolinModesO1}Noise-weighted averaged ASD showing the first harmonic violin mode region for H1 (red trace) and L1 (blue trace) for the O1 observing run.}
\end{figure}

Some of the lines observed in an amplitude spectral density of the detector data are well-understood: for example, 60~Hz power mains, mechanical resonances of mirror suspensions known as ``violin modes'' (see figure~\ref{fig:ViolinModesO1}), calibration lines, and simulated GW signals known as ``hardware injections''. Other lines are less understood and require considerable investigation to determine their nature.

The majority of instrumental lines that degrade CW searches have Q-factors in excess of $\sim$$10^3$. This is, in part, because the astrophysical sources targeted by these searches have high intrinsic Q-factors, and Doppler broadening caused the Earth's orbital velocity does not decrease the Q-factor to less than $\sim$$10^4$.

Similarly, the instrumental lines that have produced correlations between sites, degrading searches for SGWB, have also had high Q-factors. This is because the correlations are produced not by single sources affecting both of the widely-separated sites, but rather by similar sources at each site that are correlated only because they produce signals at the same, or nearly the same, frequency. Some correlated lines are due to electronic sources at each site that are set to the same frequency, controlled by a single clock (GPS), which also controls the timing of the data acquisition systems. These lines have Q-factors that are, in principle, infinite. When the frequencies are not exactly the same at each of the sites, the maximum width of the instrumental lines that can produce correlations is associated with the duration of the data segments used in the cross-correlations and the line amplitude. The typical length of Fourier-transformed data segments is 60~s long and the lowest Q-factor lines that have produced inter-site correlations are the power mains-related lines with Q-factors of $\sim$$10^3$ (the LIGO sites are on different power grids that are not synchronized).

The primary source of lines with sufficiently high Q-factors degrading both CW and SGWB analyses are processes controlled by electronic clocks or oscillators. This includes digital processes, analog electronics, and mechanical processes controlled by electronic clocks, e.g., stepper motors. Most mechanical systems do not have Q-factors above $10^3$ and so do not directly contaminate the searches by causing additional outliers, but instead degrade the sensitivity of these searches. The main exceptions are mechanical systems that are designed to have high Q-factors in order to concentrate noise in a narrow frequency band, like the ``violin'' suspension modes.

\begin{figure}[tbp]
\includegraphics[width=\columnwidth]{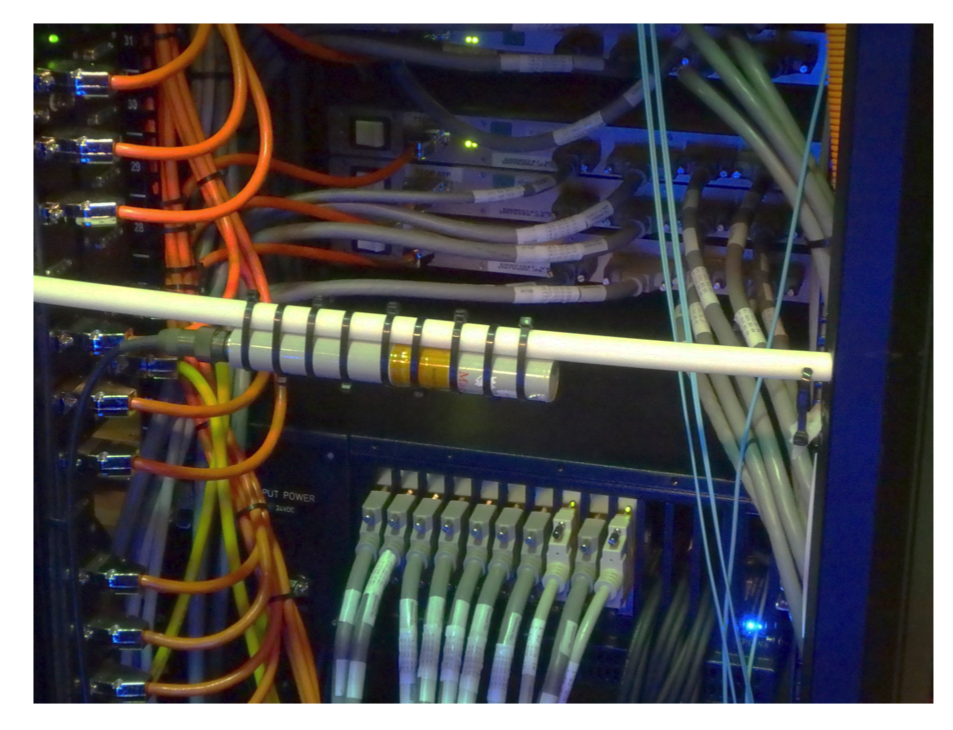}
\caption{\label{fig:Rack}Method of monitoring electronic components and cables for frequencies of instrumental lines found in the data. A Bartington fluxgate magnetometer (Mag-03 MCES100) is mounted on the horizontal white PVC pipe in the back of an electronics rack containing electronics that control the position of important optics. If the magnetometer detects fields from currents varying at the same frequency as an instrumental line, the source of the line may be in the vicinity. In addition to helping with searches for sources of line artifacts, the magnetometer can indicate that a spectral line is not astrophysical in origin.}
\end{figure}

Monitoring the frequencies associated with electronic systems is thus the main way we detect the sources of problematic instrumental lines. Monitoring each individual electronic component in the complex electronic system of LIGO would be difficult. Instead, we attempt to monitor multiple electronic systems at once, using fluxgate magnetometers (Bartington Mag-03 series, with sensitivities of about $5\times 10^{-12}$ T). The magnetometers are placed in the experimental areas and especially in important electronics racks in the electronics rooms (see figure~\ref{fig:Rack}). These magnetometers can detect even low-amplitude periodic currents controlled by oscillators and clocks that can produce high Q-factor line artifacts (detecting as low as $5\times 10^{-5}$~A at 1~m from long wires or traces).

The process of addressing lines or combs typically proceeds in three steps: identification of noise in the GW strain channel, data analysis to determine properties of the noise (precise frequency, other sensors that may witness the noise, start or end times, etc.) which may suggest a cause, and on-site investigations or interventions to mitigate the noise at its source (more details are given in section VI). This process is often iterative and experimental. Work on site is limited by available time, and also by the risk of interventions creating new problems, so noise sources are typically prioritized for follow-up by their strength, pervasiveness (number of bins contaminated), and the ease of addressing the most probable cause of the noise. Lines which are not identified or cannot be mitigated during an observing run are cataloged afterward; this is not ideal, but it does aid searches in cleaning data and rejecting outliers.

Mitigation efforts can prove challenging. In many cases, low-level spectral artifacts and combs are not visible in short-duration Fourier transforms. Only by performing averages over many days to weeks of data, do these features become obvious; hence it can take of order days to weeks of new data collection to determine if a mitigation attempt has improved the data or not. Unintended configuration changes that lead to line generation can also take time to appear, be tracked down and mitigated. As a result, significant epochs of a data run can be badly contaminated in some spectral bands, even when those bands are relatively clean at the start of the run. 

\begin{figure*}[tbp]
\centering
\subfloat[][]{\includegraphics[width=0.5\linewidth]{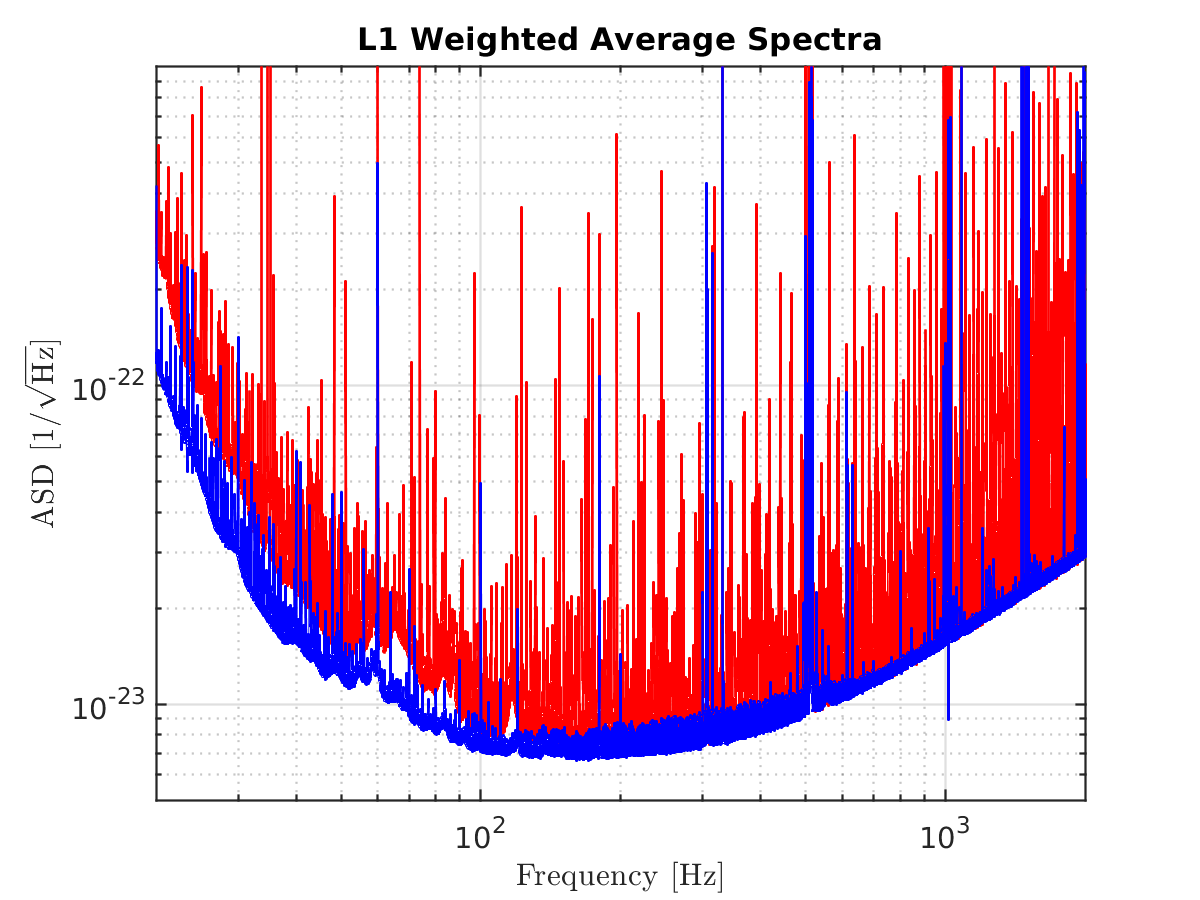}}
\subfloat[][]{\includegraphics[width=0.5\linewidth]{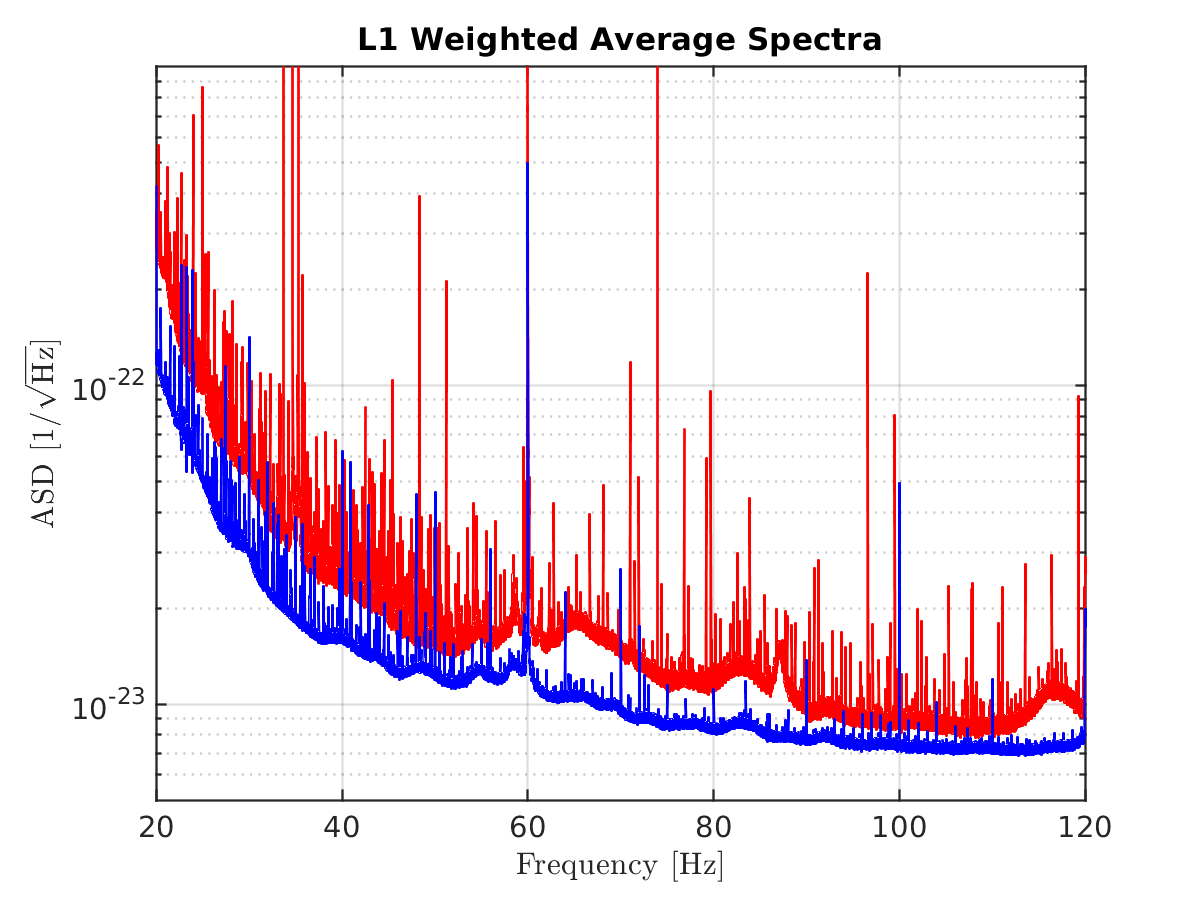}}\\
\subfloat[][]{\includegraphics[width=0.5\linewidth]{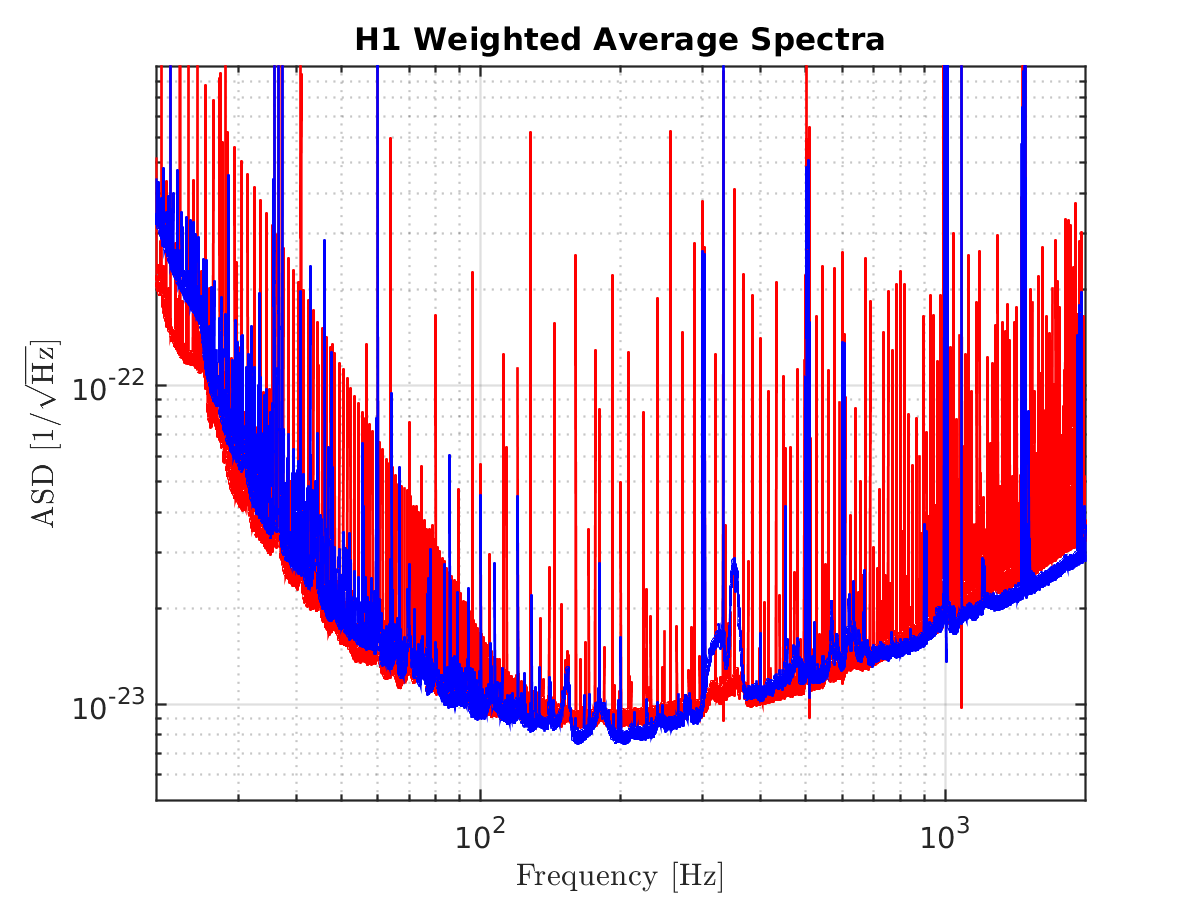}}
\subfloat[][]{\includegraphics[width=0.5\linewidth]{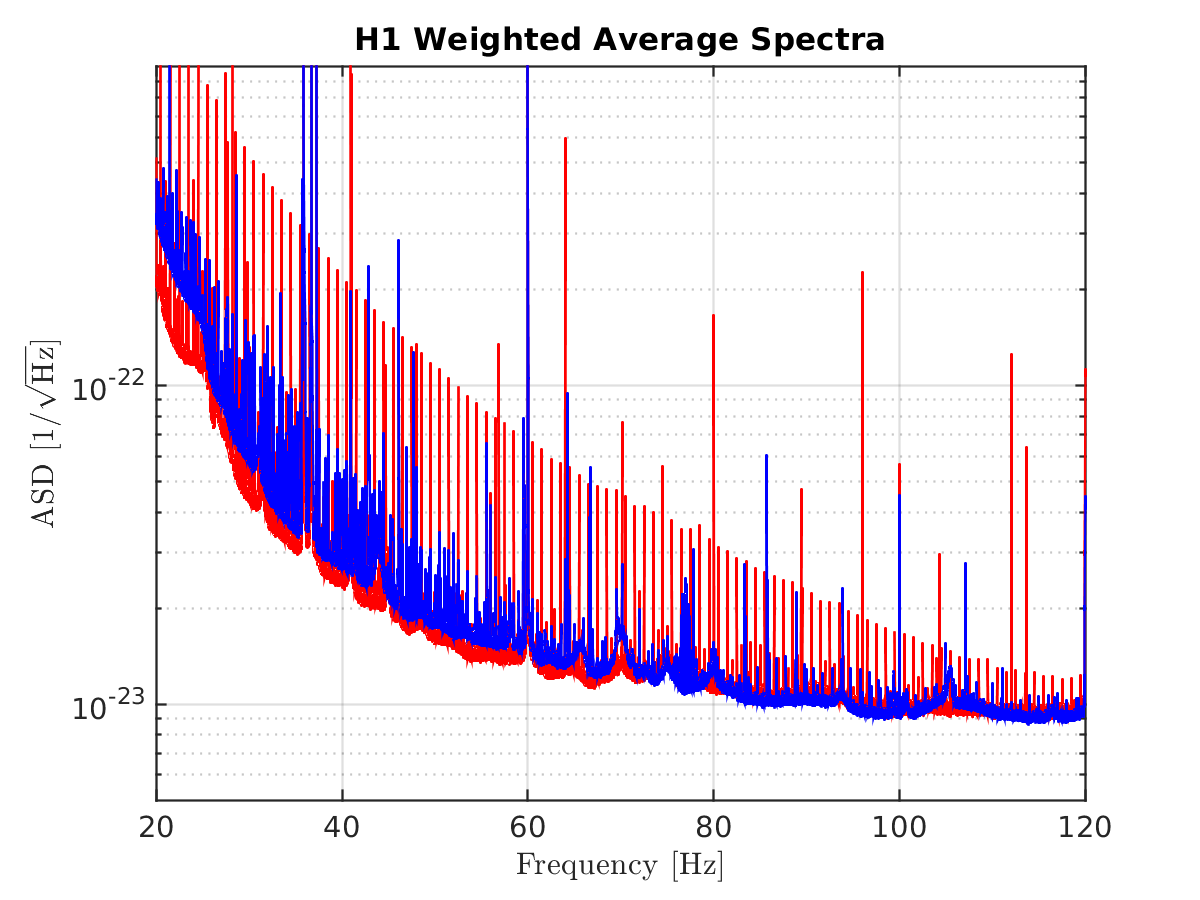}}
\caption{\label{fig:spectra}Average amplitude spectral density plots for the L1 (plots (a) and (b)) and H1 (plots (c) and (d)) detectors during O1 (red trace) and O2 (blue trace). Each individual amplitude spectral density that contributes to the average is weighted by the inverse square of its running median, so that those spectra with degraded sensitivity (higher amplitude spectral density) are de-weighted (contributing less) in the final average. (a) and (c): data in the most sensitive frequency band of the LIGO detectors 20~Hz--2~kHz. (b) and (d): data in the low frequency region from 20~Hz to 120~Hz.}
\end{figure*}

As can be seen in figure~\ref{fig:spectra}, the amplitude spectral density (ASD) of L1 and H1 exhibit different line artifacts and have somewhat different noise floors, explained in part by different configuration choices and by different environmental influences~\cite{environmental}. As a result, the couplings and the noise sources are different, and the lines and combs that need to be followed and eliminated differ between the detectors, although some common artifacts can be studied jointly. This figure also shows the improvement in data quality for long-duration searches from O1 data to O2 data, because of the investigation and mitigation activities described in section VI. We show the spectrum only between 20-2000~Hz, over which the searches for persistent GW are typically performed.

\section{\label{sec:coupling}Noise coupling mechanisms to the gravitational wave channel}

\subsection{Coupling through shared power and grounds}

Most of the mitigated lines in Initial and Advanced LIGO have coupled through shared power supplies. An electronic component draws current at a particular frequency from a power supply, which results in a small periodic drop in voltage. If a sensitive piece of electronics, such as an optic actuator driver or analog-to-digital converter, shares the power supply, the frequency can be imprinted on a signal controlling alignment of an optic, for example, and thus causes a coupling to the interferometer light. This imprinting may happen, if, for example, a gain or offset in the sensitive equipment varies with the voltage from the power supply. The solution has been to place the source of the periodic current draw on a separate power supply. This has led us to attempt to better regulate power, and to isolate noise-sensitive electronics on separate power supplies, but this is sometimes difficult to do in practice.

Coupling through shared grounds is a similar mechanism. Even when the source of the periodic current draw is on a separate power supply from the sensitive electronics, the source may affect the sensitive electronics by producing periodic voltage variation in shared grounds.

\subsection{Coupling through magnetic or electrostatic fields}

Another common coupling mechanism has been direct coupling of magnetic fields to sensitive control systems or signals. For example, we have observed fields from switching power supplies coupling magnetically to signals passing through analog-to-digital converters. We have also observed 60~Hz mains magnetic fields coupling directly to permanent magnets that are mounted on certain optics for actuation. However, in Advanced LIGO, our main magnetic coupling is through cables and connectors. Mitigation efforts have included separating cables, smaller actuation magnets, electrostatic actuation, active cancellation, reducing stray fields, and separating sources and coupling sites. Digital communication systems, such as those that use Ethernet, are a common source, but it is not always easy to keep them away from sensitive systems.

When electrostatic fields are generated inside of the vacuum chambers, they may couple directly to the test masses. Electrostatic fields may also couple to control signals at locations where shielding is imperfect, like connectors. Investigations have suggested that certain sources couple through periodically modulated electrostatic fields, although this mechanism has not been unequivocally demonstrated.

\subsection{Mechanical coupling}

Thermally-excited high Q-factor resonances of the wires suspending optics have produced problematic lines for the CW searches by vibrating the suspended optics, which causes modulation of interferometer light, and thus couples optically to the GW strain channel. The precise frequencies of secondary suspensions may not be known in advance. Most other mechanical components are low Q-factor by design, and the broad lines that they produce typically only degrade search sensitivity for CW signals. Mechanical systems that are controlled by clocks, like stepping motors or some fans, might have Q-factors that are high enough 
to be problematic, but these have not been among the sources that we have found.

\subsection{Data acquisition artifacts and non-linear coupling}

We have observed lines and combs produced by aliasing of high-frequency spectral artifacts, as well as artifacts from digital-to-analog converters. Additionally, we have observed inter-modulation products between lines of known or unknown sources during certain periods of data collection. It is also likely that we have observed combs produced by occasional errors in transmission of digitized data within the data acquisition system. The fundamental frequency of the comb is determined by the frequency (e.g., 16~Hz) of a process associated with the error.

\section{\label{sec:tools}Data analysis tools}

In this section we briefly describe some data analysis tools used to monitor and analyze the data quality for persistent gravitational wave searches.

\subsection{\label{sec:fscan}Fscan}

\begin{figure*}[tbp]
\centering
\subfloat[]{\includegraphics[width=0.5\linewidth]{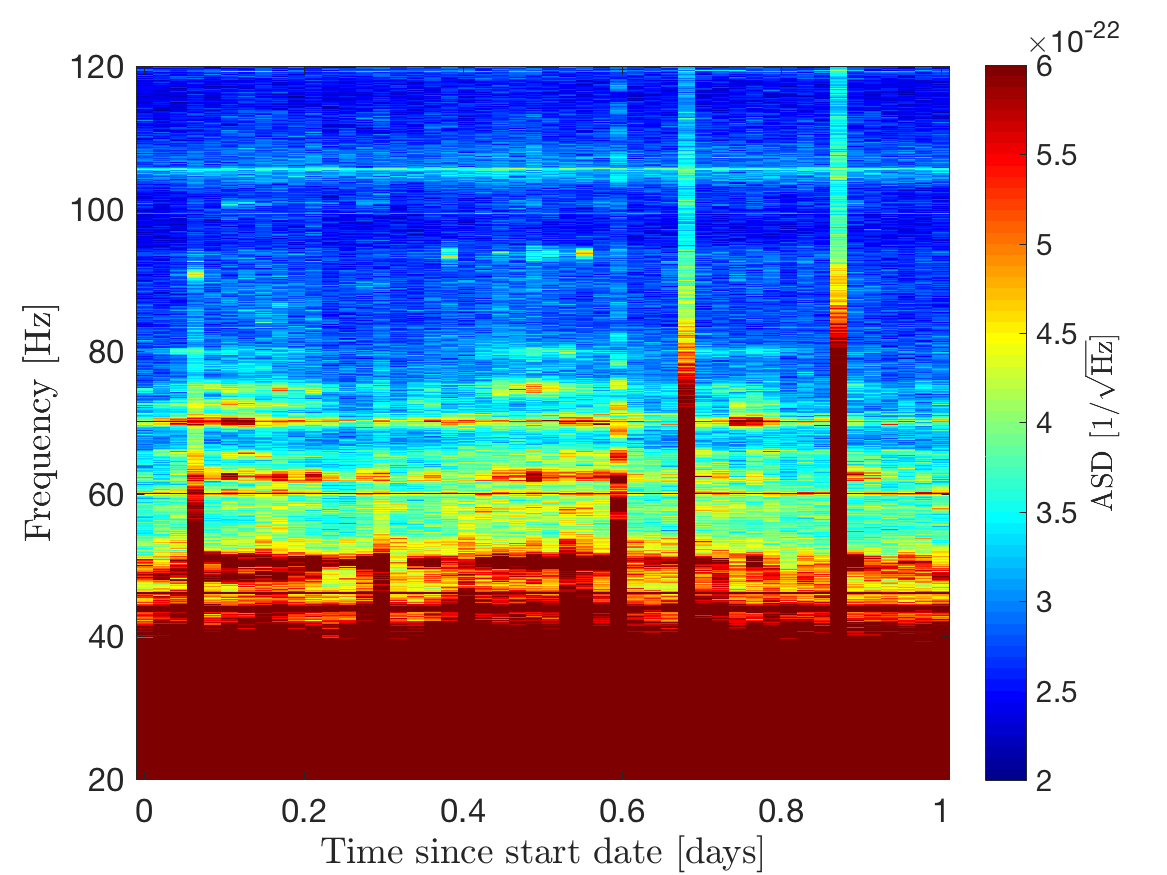}}
\subfloat[]{\includegraphics[width=0.5\linewidth]{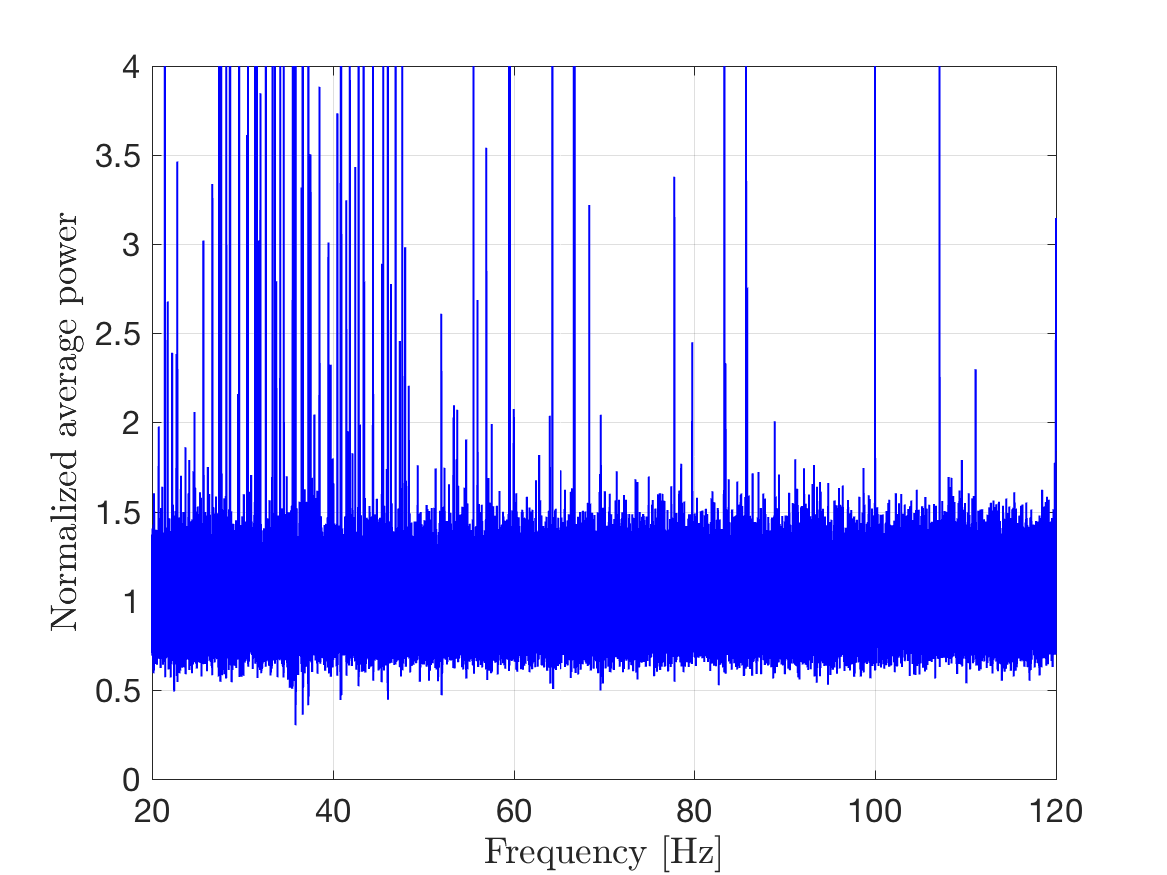}}
\caption{\label{fig:Fscan}Typical plots produced by Fscan: (a) shows a spectrogram of one day (23 April 2017) of Hanford strain data (with color-coded amplitude); (b) shows the corresponding daily averaged normalized power versus the frequency.}
\end{figure*}

Fscan is a tool that finds and monitors spectral lines~\cite{CoughlinLVC}. It uses data from the GW strain channel and hundreds of auxiliary channels for each detector, and it produces ``Short Fourier Transforms'' (SFTs) of 1800-s-long data segments. Fscan produces two different types of graphs: it averages the daily SFTs (with a maximum of 48 SFTs) to produce normalized power spectra in bands of default 100-Hz width and frequency binning of 1/1800~Hz for each channel, and it produces spectrograms with averaging of adjacent frequency bins (default bin resolution of 0.1~Hz). In the absence of non-Gaussian artifacts, the normalized spectra should be flat with random fluctuations about an expectation value of one, where the underlying statistical distribution would be $\chi^2$ with a number of degrees of freedom equal to twice the number of SFTs used to construct the spectra. Figure~\ref{fig:Fscan} shows an example of these two types of plots. Thousands of such graphs are generated automatically each day for each observatory from the GW strain channel and auxiliary channels, to provide a reference archive for line investigations.

In addition, the strain channel SFTs are used to produce (unwhitened) inverse-noise-weighted spectral averages for each day and cumulative from the start of the run through that day. The inverse noise weighting is meant to mimic the weightings used in many CW searches~\cite{CWO1AllSky}, which weight more heavily those time spans with better sensitivity. Comparing such spectral averages with arithmetic averages also allows rapid identification of non-stationary line artifacts.

\subsection{FineTooth}

\begin{figure*}[tbp]
\includegraphics[width=\linewidth]{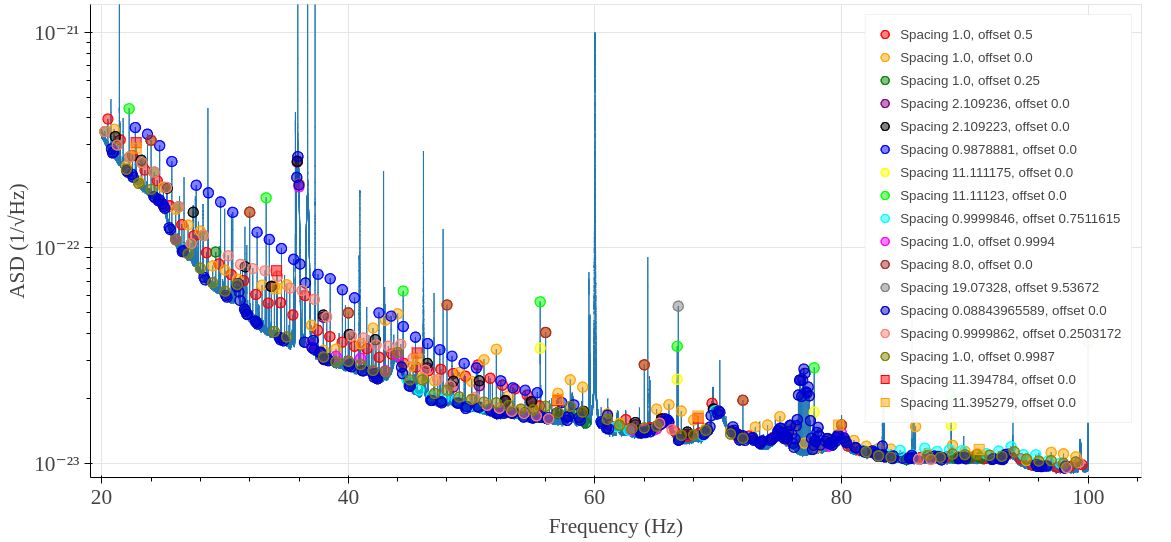}
\caption{\label{fig:FineTooth}A screenshot showing the comb plotting feature of FineTooth, on a run-averaged spectrum from Hanford in O2.}
\end{figure*}

FineTooth is a set of tools to help identify combs and monitor them over time. It is comprised of a plotting tool, a tracker for known combs, and a comb finding tool. The plotting tool creates interactive browser-based plots using the Python library Bokeh, allowing the user to overlay combs and lines and easily explore spectral features, as shown in figure~\ref{fig:FineTooth}. The tracker accepts a list of known combs and a list of channels, and then draws from Fscan data to create plots showing the historical strength of each comb in each channel. The comb finding tool searches for common spacings between peaks of comparable heights, generating a list of potential comb candidates to be vetted by the user.

During observing runs, the FineTooth tracker is run daily on a series of magnetometer channels which typically witness noise in nearby electronics, as well as on daily and run-cumulative spectra from the GW strain channel, providing a summary page for data quality checks and a tool for rapid investigation of specific combs. The comb finding and plotting tools are also used to provide an 
alert for new combs appearing in the cumulative spectrum mid-run, and to aid in comb identification for the purpose of generating vetted noise line lists.

\subsection{NoEMi}

NoEMi (Noise frequency Event Miner) is a tool used for line monitoring and as a line database~\cite{NoeMi}. It runs daily and weekly, using data from the GW strain channel and several auxiliary channels, calculating FFTs with 1 mHz resolution. It creates time-frequency diagrams from the peaks found in the spectra; the program also calculates the persistency of the lines (number of peaks in that frequency bin divided by the number of FFTs) and their critical ratio (difference between the peak amplitude and the mean value of the spectrum, divided by the spectrum standard deviation).
The persistency helps to identify loud stationary lines, while the critical ratio helps to identify non-stationary lines lacking persistency. 

NoEMi can provide the starting and end times of a line in the data. It can also follow wandering lines, by allowing some change in frequency between different time periods. NoEMi looks for coincidences (lines with the same frequency) between the GW channel and the other channels, calculating a value between 0 and 1 to quantify the probability of coincidence for each different auxiliary channel. This automated coincidence monitoring is especially valuable when searching
for causes of line artifacts seen in the GW strain channel.

\subsection{\label{ssec:coherence}Coherence}

Searches for an SGWB are done by cross-correlating the strain channel data in two detectors in the frequency domain~\cite{Christensen:1992wi,Allen_Romano_1999,sph_methods,rad_methods}. 
Depending on the source model considered, SGWB searches can be \emph{broadband}, where the signal is spread over a range of frequencies, or \emph{narrowband}, where the signal is concentrated in a narrow frequency band. Additionally, SGWB searches can either target specific sky directions using the time-delay between detectors, or integrate over a range of physical time delays assuming the source is isotropically distributed or otherwise extended on the sky.

If there is a source of noise that is coherent between the two detectors then it will show up as an excess in the cross-correlation. We must therefore cross-check our GW data streams with local environmental channels to verify that any excess in our cross-correlation statistic is not due to a local source of noise. We do this by calculating the coherence between a GW data stream and many local environmental monitoring channels. We also monitor the coherence between our two GW data streams with no phase shifts.

We define the coherence as the normalized product of the Fourier transform of two data channels, $\tilde s_1(f)$ and $\tilde s_2(f)$:
\[
C(f) = \frac{|\tilde s_1^*(f) \tilde s_2(f)|^2}{|\tilde s_1(f)|^2|\tilde s_2(f)|^2}.
\]
If the detector outputs $\tilde s_{1,2}(f_i)$ are uncorrelated Gaussian random variables, then the coherence follows an exponential distribution
\[
P(C) = N e^{-CN},
\]
where $N=T \delta f$ is the number of time segments used to compute the coherence, $T$ is the observation time, and $\delta f$ is the frequency bin width. Frequency bins with a large coherence between two detectors can be identified by looking at outliers of a histogram of coherences.

\subsubsection{Coherence between strain data of two GW detectors}
\begin{figure*}[tbp]
\begin{center}
\includegraphics[width=0.4\textwidth]{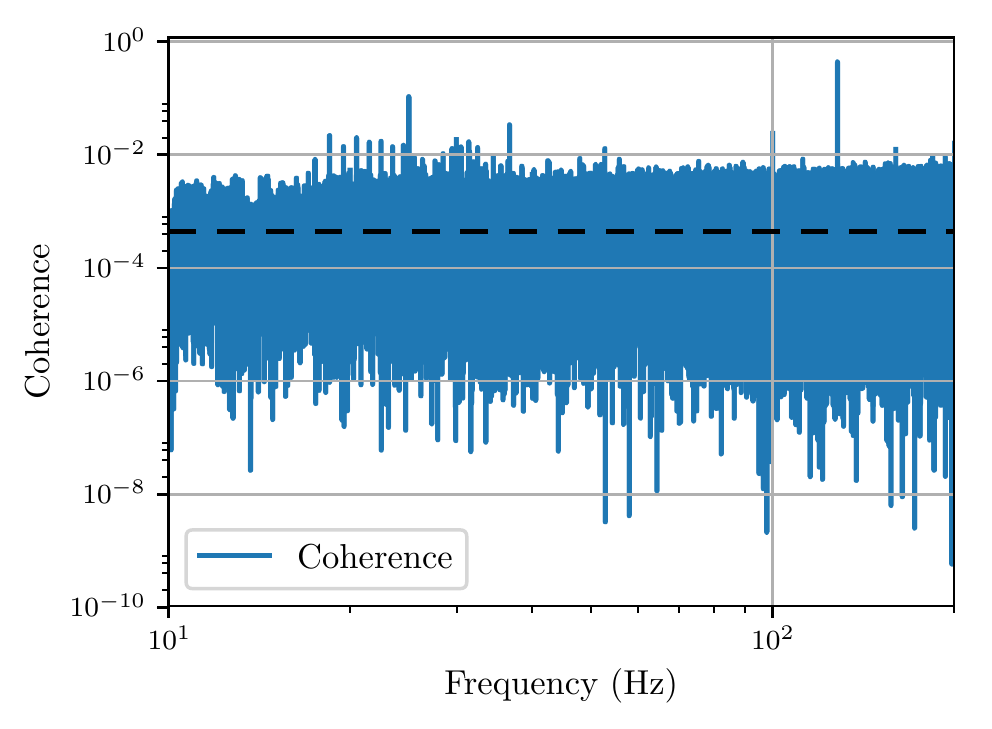}
\includegraphics[width=0.55\textwidth]{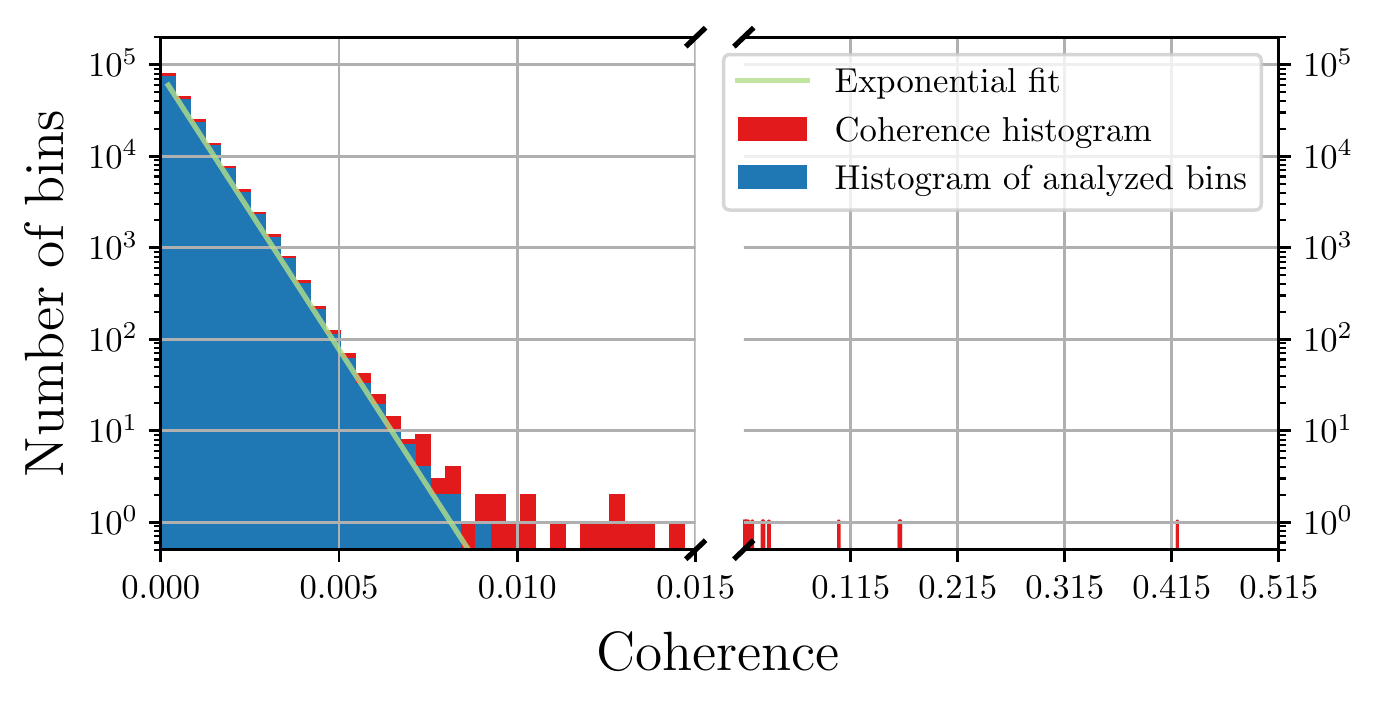}
\caption{\label{fig:H1L1_coherence}Coherent lines in O1. In the left panel, the coherence spectrum is shown between Hanford and Livingston detectors in the frequency band 10-200~Hz with 1~mHz resolution measured over the full O1 data run. The horizontal dashed line shows the expected mean value of the coherence based on uncorrelated Gaussian noise. Individual frequency bins where the coherence rises above the noise floor indicate strongly coherent lines. In the right panel, the distribution of coherences in each frequency bin is shown, compared to the behavior expected for uncorrelated Gaussian noise, in the frequency band 20-200~Hz with 1~mHz resolution. Red bins show the raw coherence. Loud lines and are followed up by studying the coherence between the GW and auxiliary channels to determine if the correlation has a terrestrial origin, as described in section~\ref{sec:make_SGWB_notchlist}. Blue bins are the resulting distribution of the frequency bins after notching lines known to have known terrestrial origin.}
\end{center}
\end{figure*}

The coherence spectrum is monitored between the two spatially separated LIGO detectors, and any excess in this spectrum at individual frequencies is followed up. Typically, we monitor the time-integrated coherence spectrum on day, week, month, and ``full run'' time-scales. This allows us to try to narrow down specific times when inter-site coherence between GW channels is higher. Any loud, narrow frequency lines is also followed up. The follow-up is done by searching for a similar excess coherence at the same frequency in the coherence spectrum of a GW strain data channel with a local environmental monitor of the same detector. Any excess coherence between a strain data channel and a local environmental monitor that is expected to be independent of the strain data channel is enough to suggest that the inter-site coherence is likely caused by a non-astrophysical source of noise. In figure~\ref{fig:H1L1_coherence}, we show a coherence spectrum made from computing the coherence over all of O1 between the Hanford and Livingston strain data channels. We also show the distribution of coherences for all 1~mHz bins in the band 20-200~Hz.

\subsubsection{Coherence between GW and local monitoring data streams}
We use a Python-based tool to compute and monitor coherence between the GW and auxiliary channels, that is essentially unchanged since the initial detector era~\cite{CoughlinLVC}. The tool computes the Fast Fourier Transforms of each strain data and auxiliary channel and stores those intermediate data products locally. It then uses those files to compute the coherence, thereby significantly reducing the I/O relative to a system which computed these files in a single step (by a factor of $\sim$$N$ where $N$ is the number of channels). A follow-up program searches for significant lines based on absolute thresholds on the coherence value as well as exceeding thresholds based by excess coherence relative to that expected from Gaussian noise. The nominal configuration uses 1024~s segments, chosen to be sensitive to lines with mHz resolution. This configuration runs automatically on a weekly basis on the data available at each detector.

\begin{figure}[tbp]
\begin{center}
\includegraphics[width=\columnwidth]{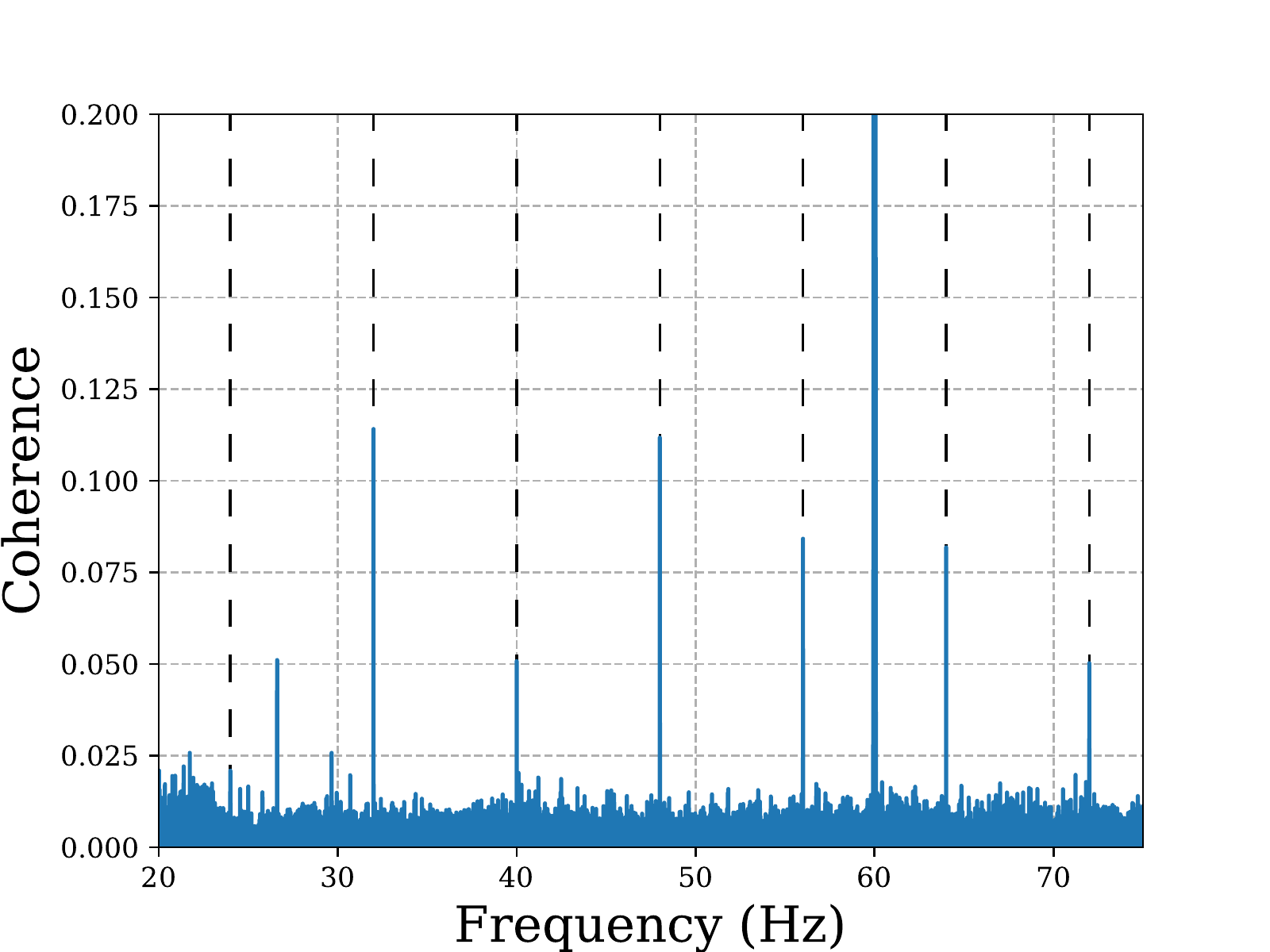}
\caption{\label{fig:coherence}Follow-up of a coherent 8~Hz comb seen in O2 using the coherence tool. The harmonics of the comb are marked with a dashed black line. The auxiliary channel used to make this plot is a monitor of the power mains at Livingston.}
\end{center}
\end{figure}

More specifically, the coherence tool used for following up inter-site GW strain channel coherence uses coherence spectra between each strain data channel and many local environment channels which have been integrated over week-long time scales. For every observing week, the coherences between the strain data channel and thousands of auxiliary channels (pertaining to the interferometer operation, as well as physical and environmental monitors) are calculated. The data used for the coherence tool has bin size of $1/1024$~Hz, and maximum frequency of 1024~Hz due to the data acquisition rate limits of the environmental monitors. To study a noise source, we need the frequencies of the noise and the resolution with which the noise is identified.  Then the tool checks the coherences from all weeks and all environmental monitors within the range of noise resolution around the frequency. If excessive coherence is found in the domain, we plot the coherence in that range for further, manual examination. If the correlation with the noise is confirmed, the channel is identified and reported. An example of coherence between a monitor of the power mains and the GW channel at Livingston is shown in figure~\ref{fig:coherence} and illustrates a visible 8~Hz comb.

\subsubsection{\label{sssec:coherence_comb_finder}Sub-threshold combs in coherence data}

For broadband stochastic searches, the final cross-correlation statistic includes an integration over frequency. While we would like to remove obvious excess coherence, there are also cases where ``sub-threshold'' combs will integrate to give a broadband excess in coherence. By this we mean there is no obvious single frequency that exceeds the typical levels of noise, but there is a set of frequencies with a specific spacing that, when summed together, gives something larger than expected if same number of bins were chosen from random noise and summed. To deal with this issue we have developed a ``comb-finder'' tool which sums the power over many possible tooth-spacings and offsets and checks whether that sum is larger than expected. 

To calculate the significance of the combined power of a set of discrete frequency bins $f_i$ representing a comb, we calculate the signal-to-noise-ratio (SNR) from the cross-correlation estimator $\hat{Y}(f_i)$ and the associated standard deviation $\sigma_Y(f_i)$ \cite{stoch_O1,stoch_s5LV}. The optimal way to combine these statistics is using a weighted sum (as described in \cite{LazzariniEA_2004,stoch_s5LV} for combining segments). For a comb with $N$ teeth, the combined statistic becomes
\begin{eqnarray}
\hat{Y}_\text{comb}^N =& \frac{\sum_i^N \hat{Y}_i \sigma_{Y_i}^{-2}}{\sum_i^N \sigma_{Y_i}^{-2}} \\
\sigma_{Y_\text{comb}}^N =& \left[\sum_i^N \sigma_{Y_i}^{-2} \right]^{-1/2}\,.
\end{eqnarray}
The subscript indicates the discrete frequency bin $f_i$ so that, for example, $\hat{Y}_i = \hat{Y}(f_i)$. 

We can define a specific comb by the offset of the first bin from the start of the search band and the frequency spacing between the teeth. The offset number of bins $m$ and spacing $n$ determine which frequency bins contribute to the comb in question. For a search over a given frequency band $\Delta f = f_\text{max} - f_\text{min}$, with a frequency resolution of $df$, the number of teeth in a comb with bin spacing $n$ will be given by $
N = 1+\texttt{floor}\left[\frac{\Delta f}{n}\right]$. We define the combined SNR statistic as
\begin{eqnarray}
\mathcal{S}_{m,n} = \frac{\hat{Y}_\text{comb}^{(m,n)}}{\sigma_{Y_\text{comb}}^{(m,n)}}
 = \frac{\sum_i^N \hat{Y}(f_{m+ni}) \sigma_{Y}^{-2}(f_{m+ni})}{\left[\sum_i^{N} \sigma_{Y}^{-2}(f_{m+ni})\right]^{1/2}}\,.
\end{eqnarray}

\begin{figure}[tbp]
\includegraphics[width=1.0\columnwidth]{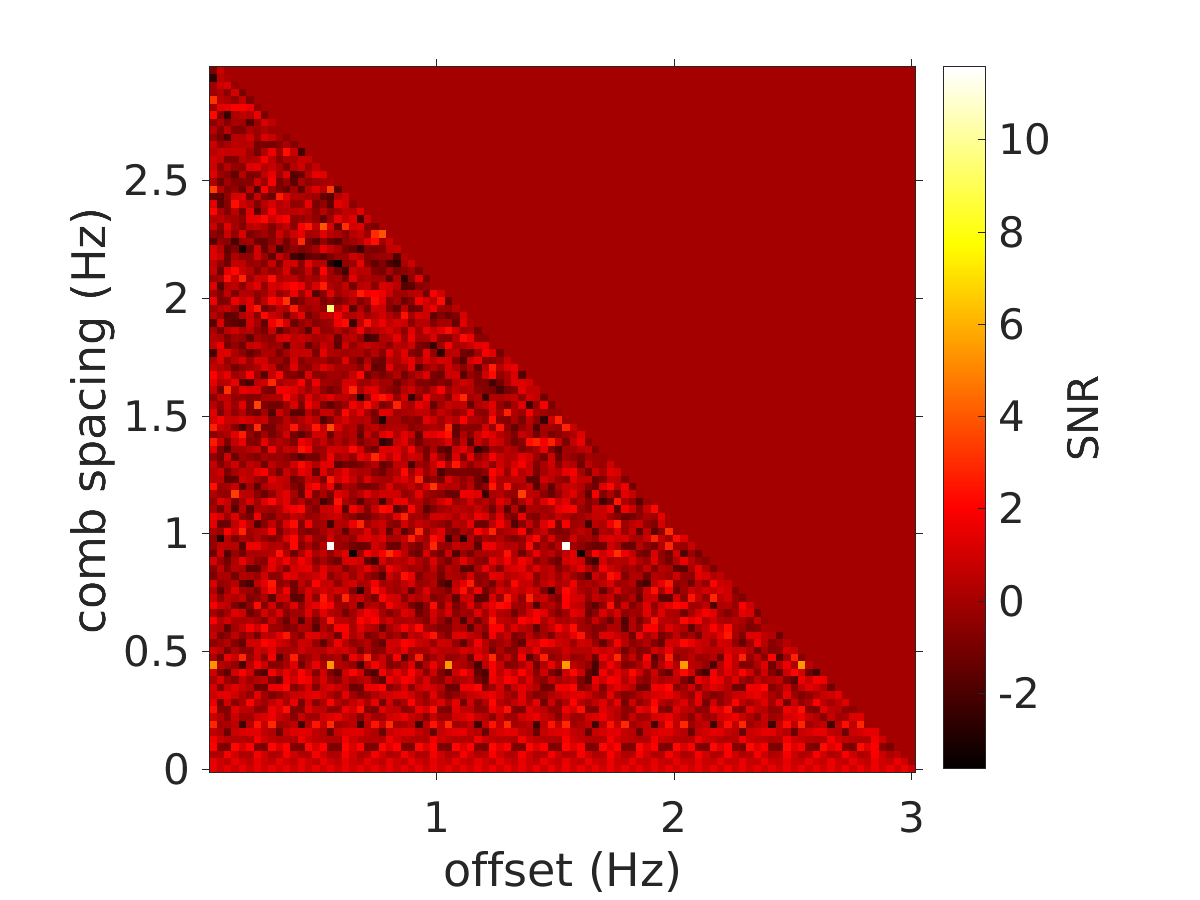}
\caption{\label{fig:O1comb}Example output of the comb-finder. White pixels indicate strong SNR. The loudest pixels indicate a coherent 1~Hz comb with 0.5~Hz offset identified during O1.}
\end{figure}

Figure~\ref{fig:O1comb} shows an example output of the comb-finder tool demonstrating the 1~Hz comb found during O1. Excess SNR is visible at regular 1~Hz spacings and offsets of 0.5~Hz.

\subsection{Folding studies}
Most line investigations are carried out in the frequency domain, but a tool has also been developed to look directly at periodicity in time-domain data, since some spectral combs arise from periodic transient glitches. The folding tool splits a long segment of data into short segments (typically a few seconds in length, corresponding to some periodicity of interest, e.g. $1/\delta f$ or $1/f_{o}$ for a comb) and averages the segments together to produce a summary plot. The data folding tool can generate daily, monthly, and full-run plots, with or without a band-pass filter applied. Band-pass filtering often makes periodicity more easily visible.

Folded data can reveal features of the periodic structures underlying spectral combs, making it useful for spotting changes that may not be evident in the spectrum. It is typically most effective for magnetometer channels (see, for example, figure~\ref{fig:FoldedMagnetometerData}), where periodicity is stronger than in the GW data channel, but on occasion periodic transients have been visible in the GW strain channel as well, most notably from
blinking light emitting diodes (LEDs), as discussed below.

\section{\label{sec:examples}Results}

In this section, we describe examples of particular noise sources that were mitigated between the O1 and O2 data runs, or during the O2 run. For each noise source, a plot showing the improvement of the spectrum in the respective frequencies is also presented.

When a new feature in the detector strain data channel is discovered by using the tools mentioned in the previous section, additional investigations to identify the source of the noise are performed:
\begin{enumerate}
\item Determine the Q-factor of the line affecting the search. This helps identify the source and type of equipment that is producing the line. If the Q-factor is above $10^6$, the source is likely to be precision-clocked electronics components, or equipment that is synchronized to GPS. Typical inexpensive clock chips in electronic devices have Q-factors of $\sim$$10^5$, though the Q-factors of newer inexpensive chips may be higher. Lines from equipment using 60~Hz timing from the mains, have Q-factors of roughly $10^3$. The Q-factors of LIGO suspension wire resonances vary, but many of the secondary optics are in the range of a few $\times$ $10^5$.
\item Identify and investigate any transitions in line amplitudes. If there are sudden changes in amplitude of the lines, it is often helpful to examine instrument logs for correlated changes in instrumentation or software.
\item Search for lines of the same frequency in the fixed magnetometer signals. If the line is found, it may help localize the cause. However, even if the frequency detected by the magnetometer may match the instrumental line, it may not be the cause. The probability of incorrect attribution is higher for lines that are at integer frequency values and are synchronized to GPS.
\item For lines that are detected in magnetometer channels, the location of the source can often be further narrowed by moving around a portable magnetometer to maximize the line amplitude.
\item Search for lines in auxiliary channels, especially error signals for secondary optical cavities. The lines for many secondary optic suspensions will have higher signal-to-noise ratios in auxiliary channels than in the GW data channel.
\item Search for LEDs flashing at the frequency of the lines. The periodic current drawn for the LED may cause the coupling by modulating power supply or ground voltages.
\item Temporarily shut down equipment in the candidate area, when possible, as a test. This is especially helpful if a line is stronger in a magnetometer signal than in the interferometer signal because the magnetometer can be used to more rapidly evaluate the effect of shutting down the equipment.
\end{enumerate}
	
\subsection{1~Hz with 0.5~Hz offset comb (Blinking LEDs)}
A strong comb with 1~Hz spacing and 0.5~Hz offset was observed throughout the O1 run. Initial tests showed coherence between the strain channel and several magnetometer channels in the electronics bays at the corner station and at the end of both arms. Follow-up studies using portable magnetometers found that the comb was loud around nearly all electronics, but particularly near equipment associated with the timing system. 

\begin{figure}[tbp]
\begin{center}
\includegraphics[width=1.05\linewidth]{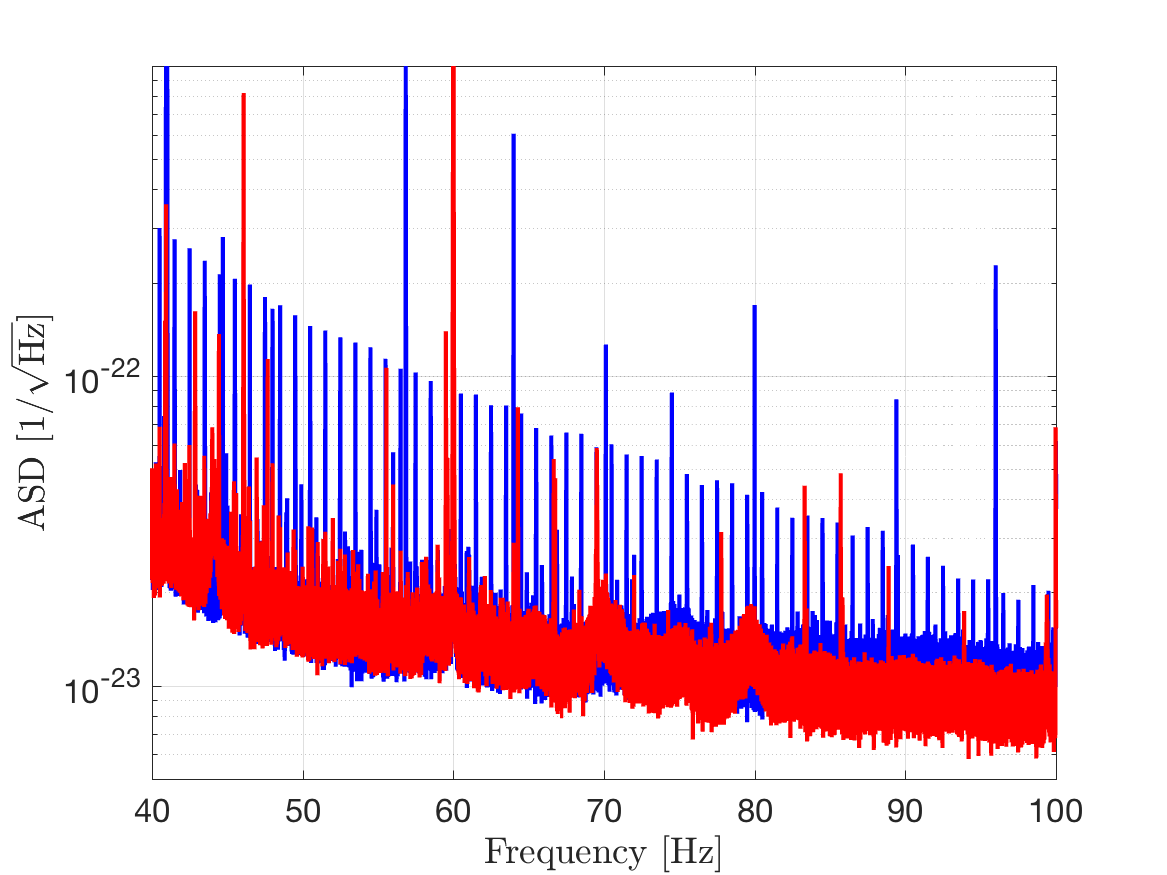}
\caption{\label{fig:HalfHzComb}Comparison of noise-weighted averaged ASD using H1 data from 5 January 2016 to 10 January 2016 (blue trace, blinking LEDs on) with noise-weighted averaged ASD from 12 December 2016 to 17 December 2016 (red trace, blinking LEDs off). The 1~Hz comb with 0.5~Hz offset is clearly reduced in the second period.}
\end{center}
\end{figure}

\begin{figure*}[tbp]
\begin{center}
\includegraphics[clip, trim=1cm 6cm 1cm 6cm, width=0.49\linewidth]{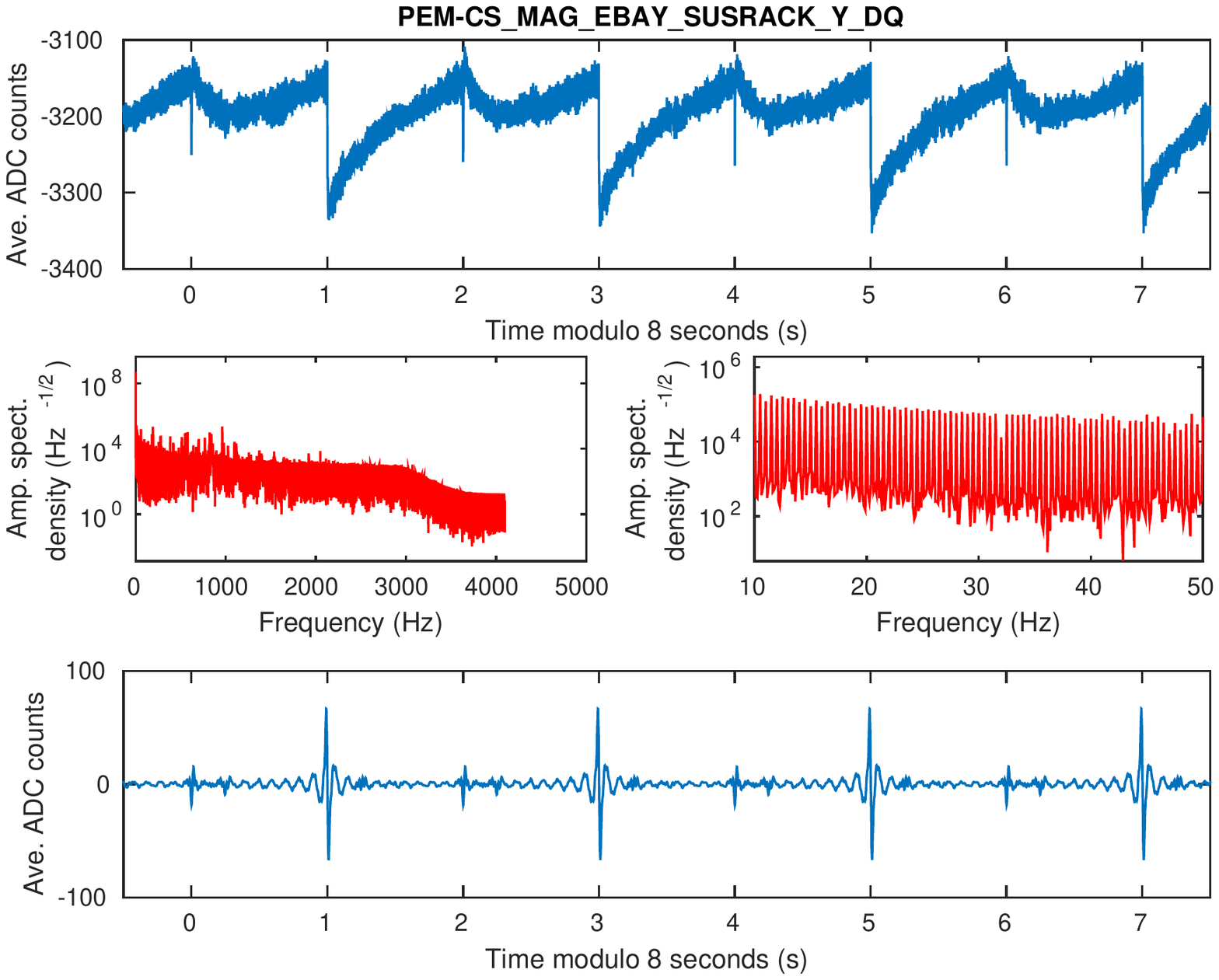}
\includegraphics[clip, trim=1cm 6cm 1cm 6cm, width=0.49\linewidth]{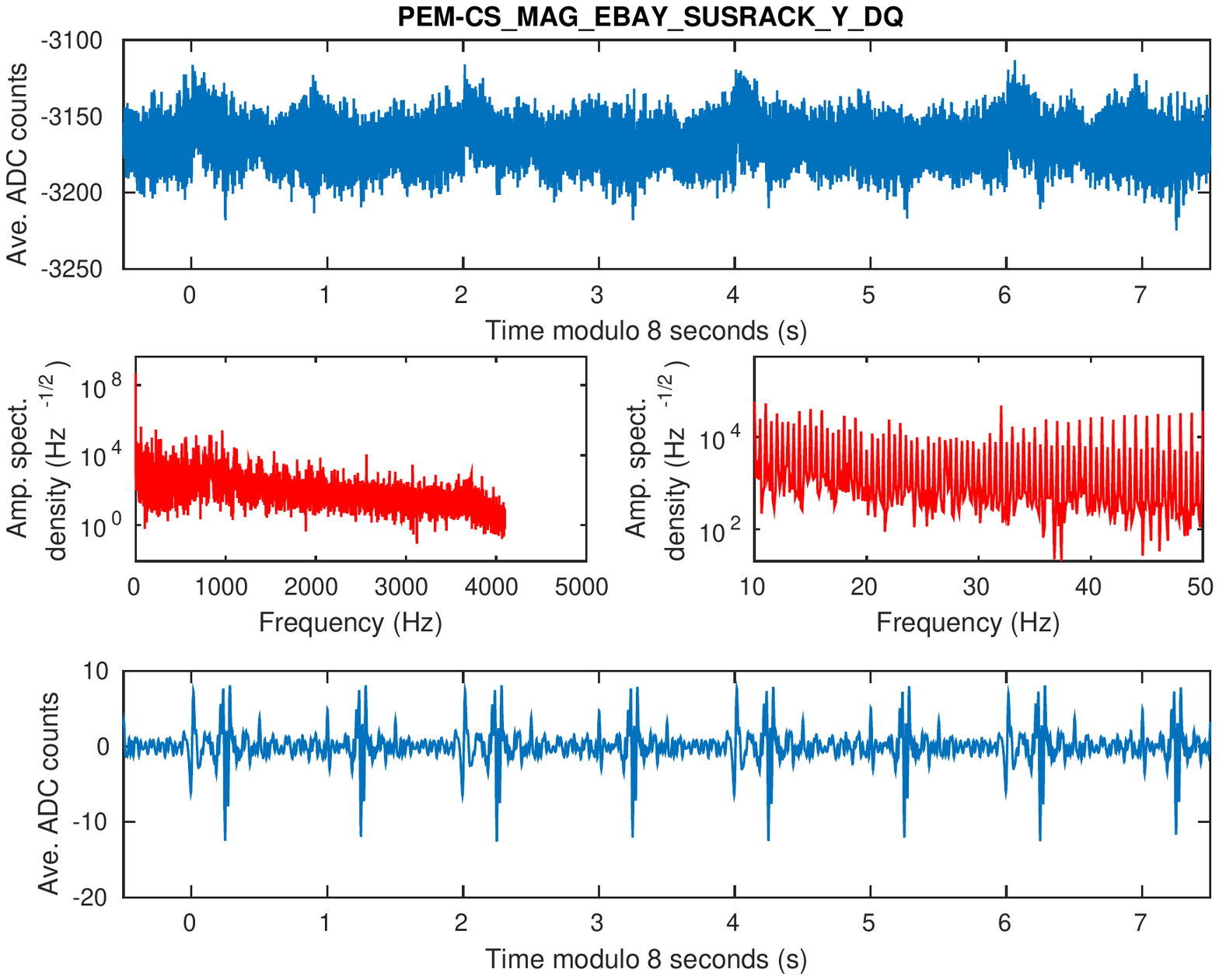}
\caption{\label{fig:FoldedMagnetometerData}Comparison of folded magnetometer data (arbitrary channel with ADC count units) between one month in the O1 run and one month in the O2 run. In each panel, the top graph shows the averaged data from the folding of 8~s intervals, the middle graphs show the power spectrum from the top graph, full-band on left and 10-50~Hz on the right (relevant to interferometer contamination), and the bottom graph is the inverse Fourier transform of the spectral data after removing bins outside of 10-50~Hz. The left panel shows results for October 2015 (O1), and the right panel shows results for December 2016 (O2). The magnitude of the glitches with 2~s periodicity is greatly reduced on the right, following mitigation of blinking LEDs after the O1 run.}
\end{center}
\end{figure*}

The master and slave components of this timing system have LED indicators that draw power in a 2~s period square wave, which would produce a Fourier series consistent with the observed comb. The slave cards were first placed on separate power supplies at both end stations and the corner station. This action did not, however, reduce the strength of the comb in the strain channel. Instead of replacing the power supply for the master system, a different approach was taken, and the firmware was updated to stop the LEDs from flashing. Shortly after this change, folding studies showed improvements in the 1~Hz periodic structures in magnetometer channels. Subsequent longer-term studies showed the change was successful in reducing the comb strength by a factor of about 10, as shown in figure~\ref{fig:HalfHzComb}. Another measure of mitigation can be seen in a comparison of folded data for a particular (arbitrarily chosen) magnetometer channel at LIGO Hanford Observatory between one month in the O1 data run and one month in the O2 data run, as shown in figure~\ref{fig:FoldedMagnetometerData}; the transients with 2~s periodicity are greatly reduced in magnitude (but not eliminated).

\subsection{8~Hz / 16~Hz comb (OMC length dither)}
The Output Mode Cleaner (OMC) is an optical cavity used to clean the recombined light that returns from the arms of the interferometer. The length of this cavity is controlled by two piezo-electric transducers, which adjust the length of the cavity and ``dither'' (modulate) this length with a given frequency. The power observed in the photodiodes is proportional to the square of the cavity length variation, which is proportional to different up-conversion and down-conversion factors coming from the beating of different noise lines and the dither line. 

During the O1 run, a strong and pervasive 8-Hz comb was observed in Hanford strain data, with especially strong even harmonics, making it appear to be a 16~Hz comb in much of the detection band~\cite{Regression}. In February 2016, following the O1 run, the frequency of the OMC dither line was changed in order to see if the spacing of the observed 8~Hz / 16~Hz comb would change~\cite{OMCDither}. When the frequency was changed from 4100~Hz to 4100.21~Hz, the dominant comb changed from a 16~Hz spacing to a 16.84~Hz spacing, consistent with a dependence on the difference between the dither frequency and 4096~Hz (1/4 of channel sampling frequency). The cause of this beat was traced down to a too-coarse digitization used in the digital-to-analog converter (DAC) for the dither actuation. Increasing the digital input by a factor of 100 and applying a compensating analog 100$\times$ suppression eliminated the comb. 

A seemingly different but related phenomenon was observed in the O1 Livingston data, namely a strong and pervasive pattern of lines with frequencies composed of integer combinations (positive or negative coefficients) of 22.7~Hz and 25.6~Hz. Following the O1 run, a test to reduce the OMC length dither amplitude by a factor of two greatly reduced the 22.7 / 25.6~Hz lines. Further investigation uncovered another non-optimum DAC input configuration for the 4800.1~Hz dither. Fixing the digitization choice eliminated the lines.

Since the non-optimized digitization for the OMC length dither for both interferometers created lines that contaminated the entire 2~kHz spectra shown in figure~\ref{fig:spectra}, the dominant difference between the O1 and O2 spectral lines seen in the left panels of the figure is due to the mitigation of the dither-induced lines. Most of the other mitigations described here mainly affected the right panels of figure~\ref{fig:spectra}.

\subsection{11.111~Hz comb (Vacuum sensors)}
A 11.111~Hz comb was found at the beginning of May 2017 in the Hanford O2 data. After some investigations with a portable magnetometer, it was found that this comb was present around cables from the 24~V power supply that powered one of the Electrostatic Drives (ESD), which control the test mass positions and so are one of the most sensitive components in the system. The components powered by the cables from this supply were checked, and a strong 11.111~Hz magnetic field was detected near a vacuum sensor. 

\begin{figure*}[tbp]
\centering
\subfloat[][]{\includegraphics[width=0.5\linewidth]{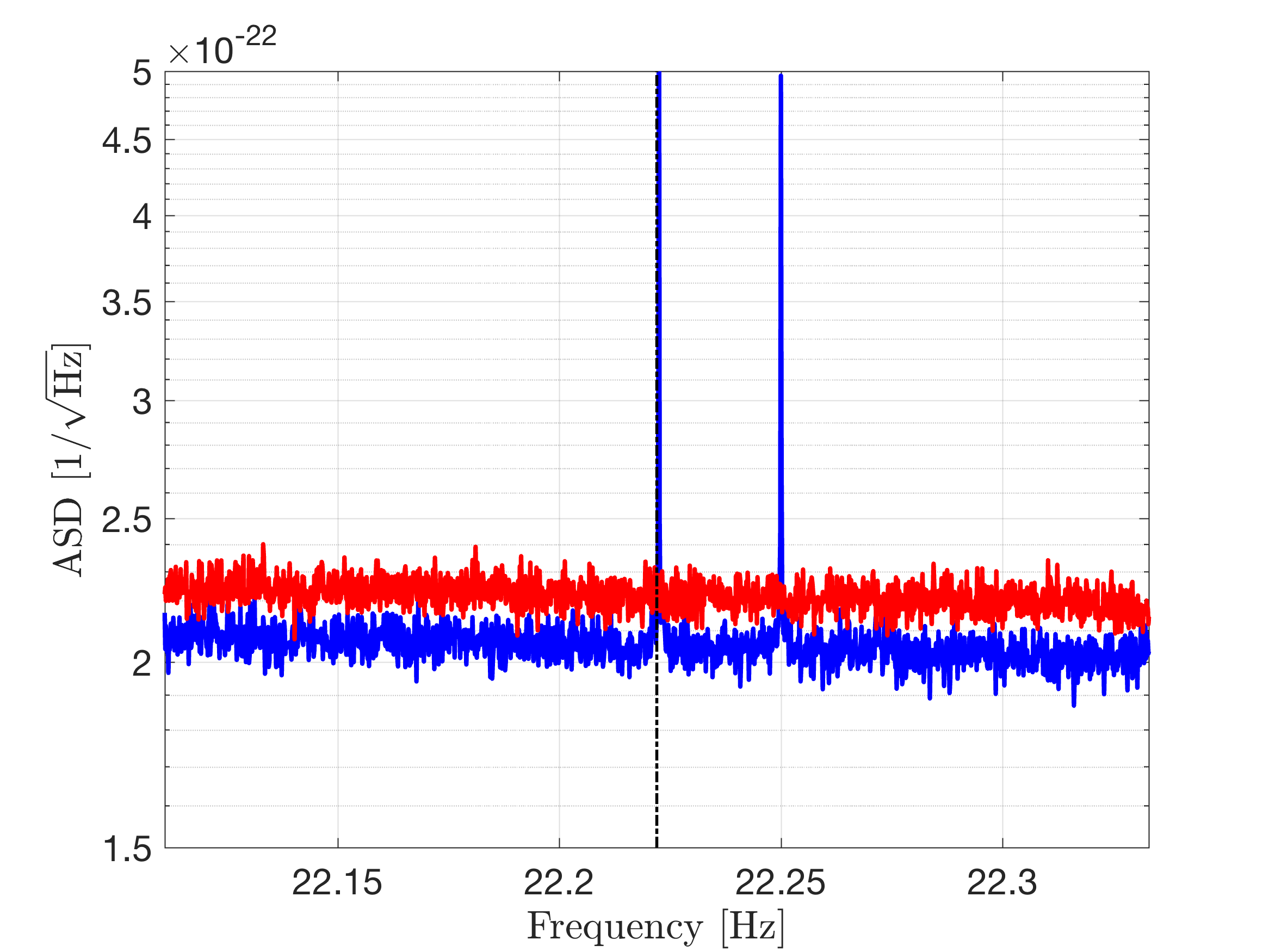}}
\subfloat[][]{\includegraphics[width=0.5\linewidth]{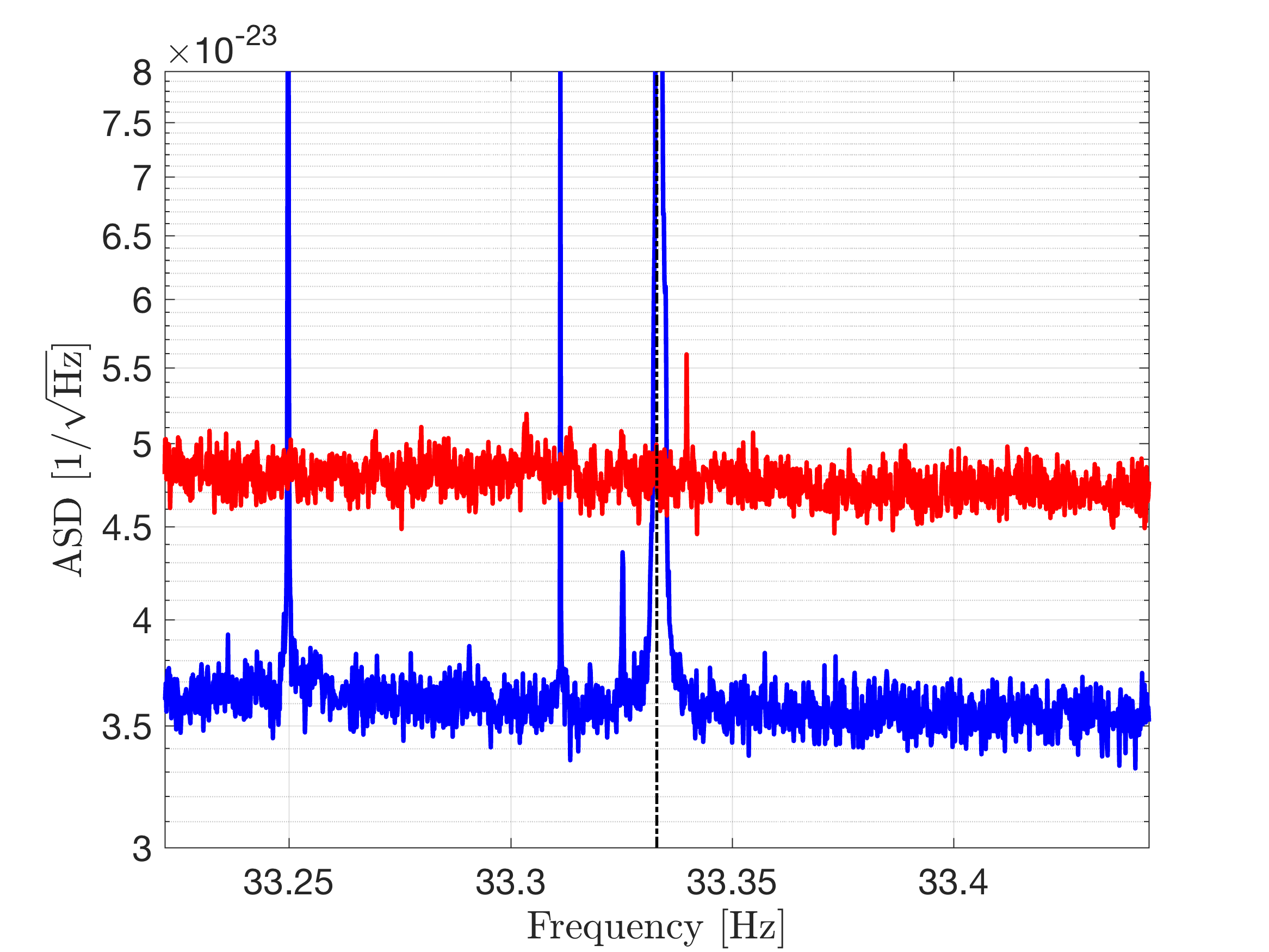}}\\
\subfloat[][]{\includegraphics[width=0.5\linewidth]{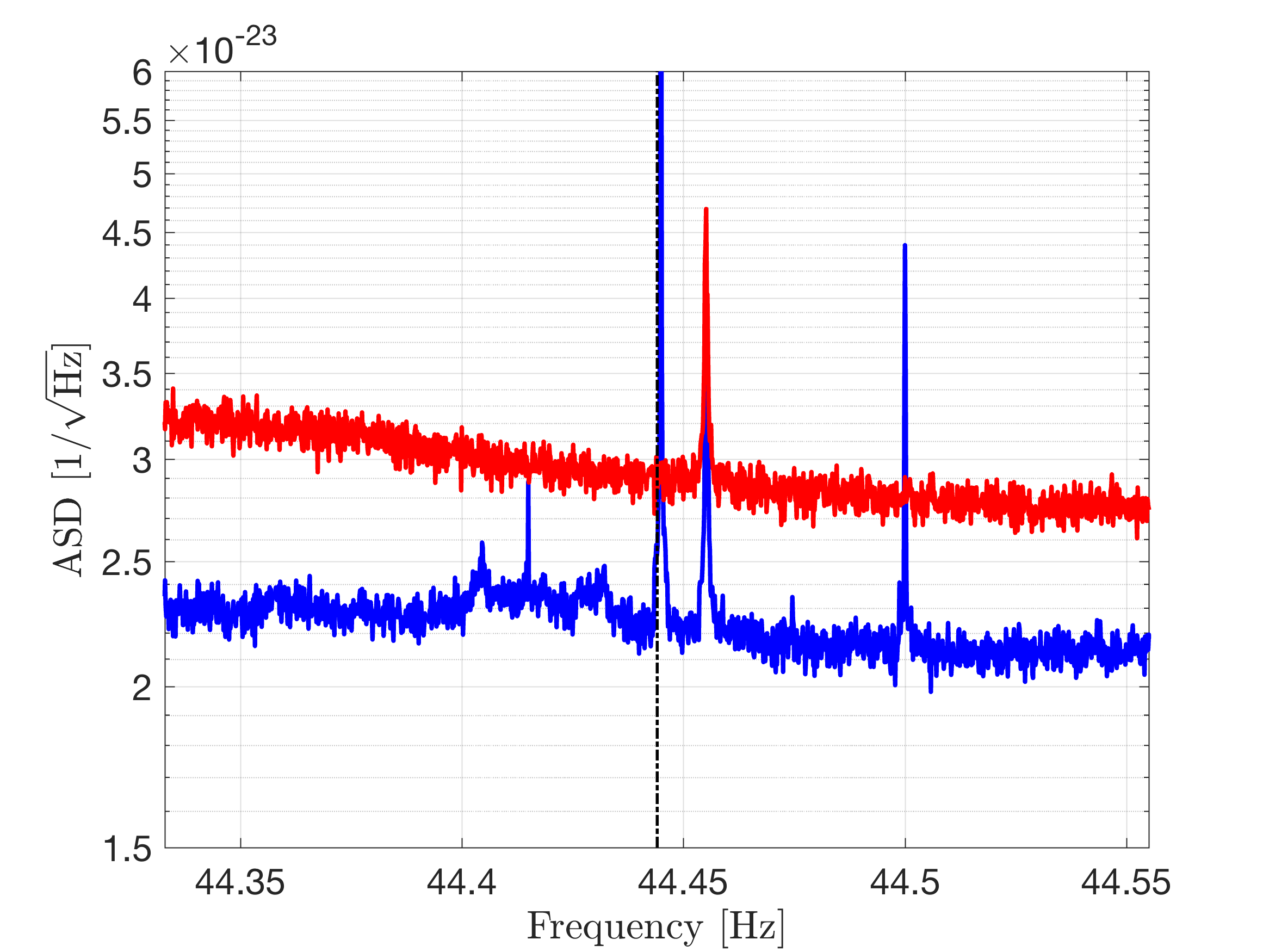}}
\subfloat[][]{\includegraphics[width=0.5\linewidth]{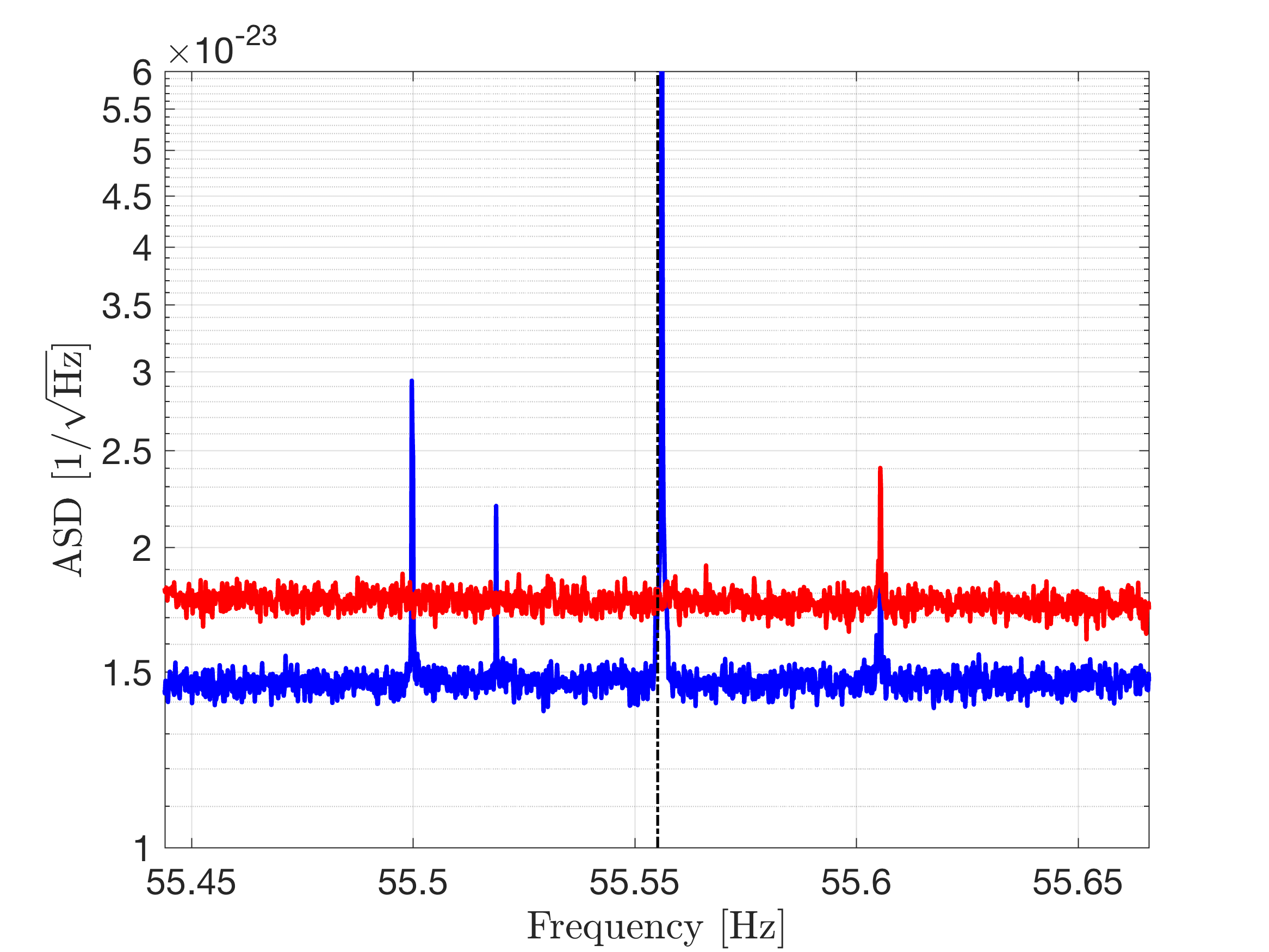}}
\caption{\label{fig:VacuumSensorsComb}Comparison of noise-weighted averaged ASD using H1 data from 8 March 2017 to 8 May 2017 (blue trace, vacuum sensors sharing ESD power supplies) with noise-weighted averaged ASD from 8 June 2017 to 25 August 2017 (red trace, vacuum sensors not sharing ESD power supplies). The 11.111~Hz comb is gone in the second period, as we can see when we look at the harmonics 2, 3, 4 and 5, shown in the four panels by a black dashed line.}
\end{figure*}

A laboratory test confirmed that the communication frequency between this type of sensor and its computer controller was 11.111~Hz, and that the LED on the sensor flashed at this frequency. The other vacuum sensors at this station were powered by separate supplies but this sensor had been connected to the ESD power supply in error. Placing the sensor on the proper power supply eliminated the comb from the GW strain channel as shown in figure~\ref{fig:VacuumSensorsComb}.

\subsection{Near-2~Hz with 1~Hz offset comb (CPS timing fanout)}
A strong comb with near-2~Hz spacing was first noted in the Hanford GW strain channel during the O1 run, and seen again during the engineering run preceding O2. High-resolution spectra were used to measure the spacing of the comb more accurately, to 1.999951~Hz, with teeth visible on odd-integer multiples from $\sim$9 to $\sim$175~Hz. It was not possible to identify from the GW strain channel alone the date on which the comb first appeared because the detector was offline for an extended period in the spring of 2016. 

Fortunately, the same comb was clearly visible in a magnetometer channel at the End X station, which \emph{was} collecting data during the detector downtime. Magnetometer data showed the comb appearing on 14 March 2016. Detector logs showed that work was done at End X on that same date: specifically, on the capacitive position sensor (CPS) interface chassis. The CPS interface chassis was in the same electronics rack as the electrostatic drive (ESD), with which it shared a power supply. Since the ESD drives the end test mass directly, this provided a likely coupling mechanism for the comb.

\begin{figure}[tbp]
\includegraphics[width=1.05\columnwidth]{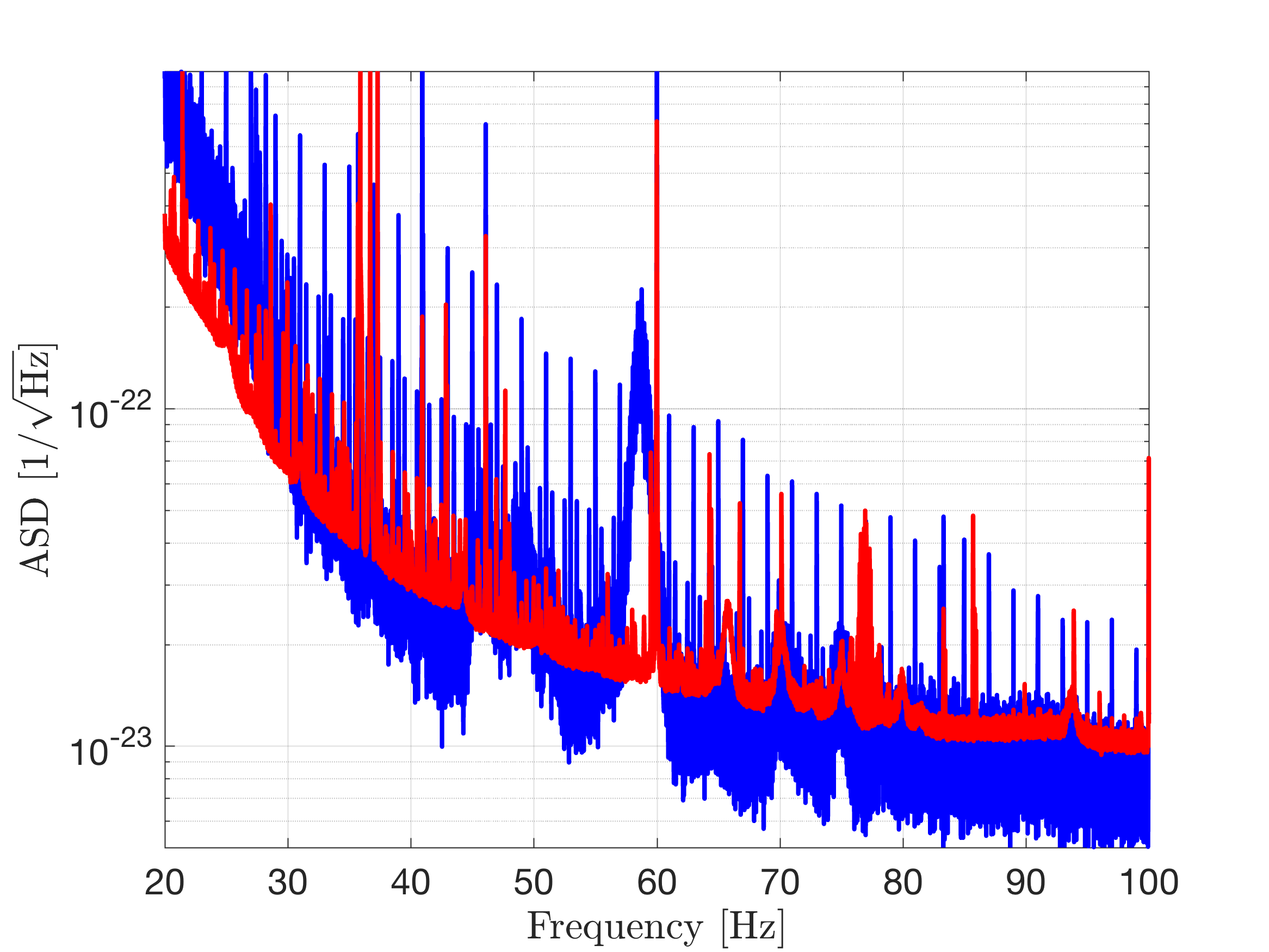}
\caption{\label{fig:CPS}Comparison of noise-weighted averaged ASD using L1 data from 14 May 2016 (blue trace, CPS timing fanout on) with noise-weighted averaged ASD from 8 June 2017 to 25 August 2017 (red trace, CPS timing fanout off). The $\sim2$~Hz comb with 1~Hz offset is mitigated in the second period.}
\end{figure}

Coincidentally, a temporary magnetometer had recently been placed near the ESD, as part of a transient glitch investigation. This magnetometer showed the comb even more clearly than the permanent magnetometer initially used for tracking. This provided solid evidence for the physical location of the coupling. The CPS timing fanout was subsequently reprogrammed, and powered on an alternate power supply, one isolated from the ESD, which mitigated the comb, as can be seen in figure~\ref{fig:CPS}.

\subsection{0.5~Hz / 2.24~Hz comb (remote control chassis)}

\begin{figure}[tbp]
\includegraphics[width=1.05\columnwidth]{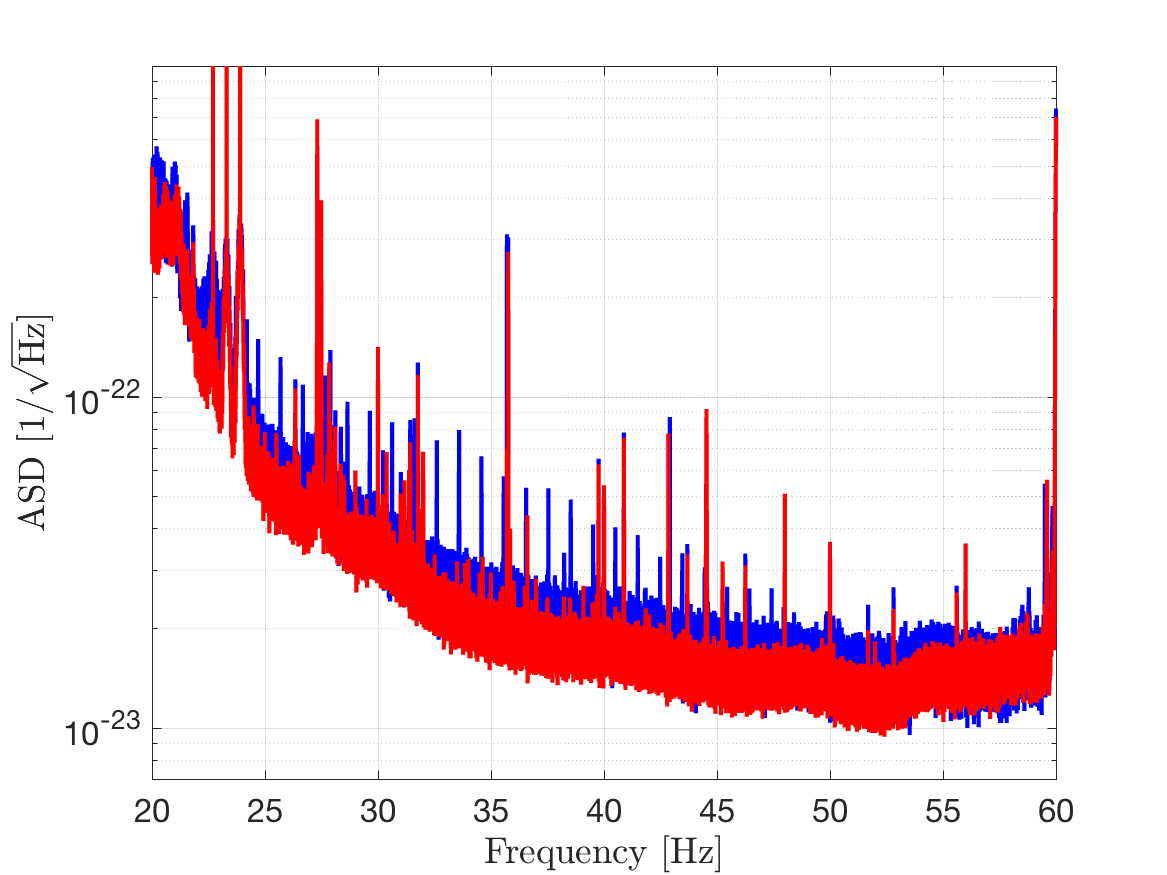}
\caption{\label{fig:IlluminatorComb}Comparison of noise-weighted averaged ASD using L1 data from 23 January 2017 to 30 January 2017 (blue trace, remote control chassis on) with noise-weighted averaged ASD from 1 February 2017 to 8 February 2017 (red trace, remote control chassis off). The 0.5~Hz / 2.24~Hz comb is attenuated in the second period.}
\end{figure}

A pervasive comb in Livingston strain data was observed throughout early O2 with two spacings: one near 0.5~Hz and the other near 2.24~Hz \cite{IlluminatorComb}. Magnetometer data indicated the comb was associated with controller chassis used for remote control operations of equipment. In particular, the controller turns on and off an illuminator used in the vacuum chamber. While this illuminator is off during normal operations, it was found that disconnecting the Ethernet and power cables from the remote-control chassis mitigated the comb. For the remainder of O2, the illuminator control remained disconnected. Figure~\ref{fig:IlluminatorComb} compares two different periods, showing the improvement in the amplitude spectral density when the illuminator was turned off. The particular coupling mechanism between the remote-control chassis and the GW channel was not determined, but a similar system at Hanford did not appear to couple.

\subsection{1~Hz with 0.25 and 0.5~Hz offsets comb (digital camera Ethernet adapter)}
\begin{figure}[tbp]
\includegraphics[width=1.05\columnwidth]{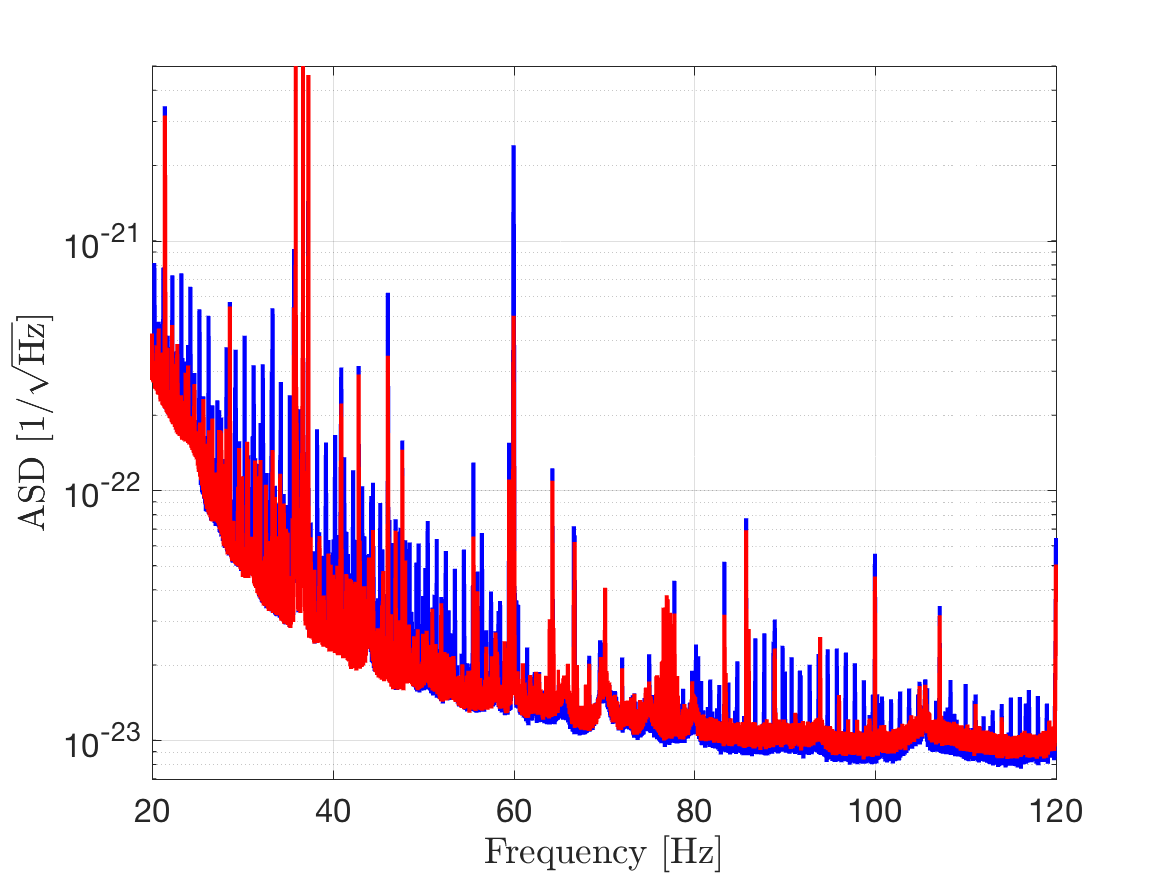}
\caption{\label{fig:EthernetComb}Comparison of noise-weighted averaged ASD using H1 data from 14 March 2017 to 18 April 2017 (blue trace, Ethernet camera adapter on) with noise-weighted averaged ASD from 19 April 2017 to 20 June 2017 (red trace, Ethernet camera adapter off). The 1~Hz comb with 0.5 and 0.25~Hz offsets is gone in the second period.}
\end{figure}

Digital cameras are mounted on the vacuum enclosures near glass view-ports to allow for imaging of in-vacuum interferometer end mirrors. These cameras can be operated remotely using a network Ethernet adapter connection. Normally, these adapters remain off during normal operations. From 14 March 2017 through 18 April 2017, however, these were inadvertently left on after routine maintenance activities~\cite{Ethernet}. Unfortunately, with these adapters on, detector data was found to have low-level, but nevertheless detrimental contamination for CW and stochastic searches. After turning off these Ethernet adapters, a mitigation of observed 1~Hz combs with 0.25~Hz and 0.5~Hz offsets was achieved, as can be seen in figure~\ref{fig:EthernetComb}. While the coupling mechanism is not certain, the possibilities include cross talk between cables and modulation of grounds. In any case, we believe it illustrates the dangers of digital signals near sensitive systems.

\subsection{86~Hz line (Pcal high frequency injections)}
\begin{figure}[tbp]
\includegraphics[width=1.05\columnwidth]{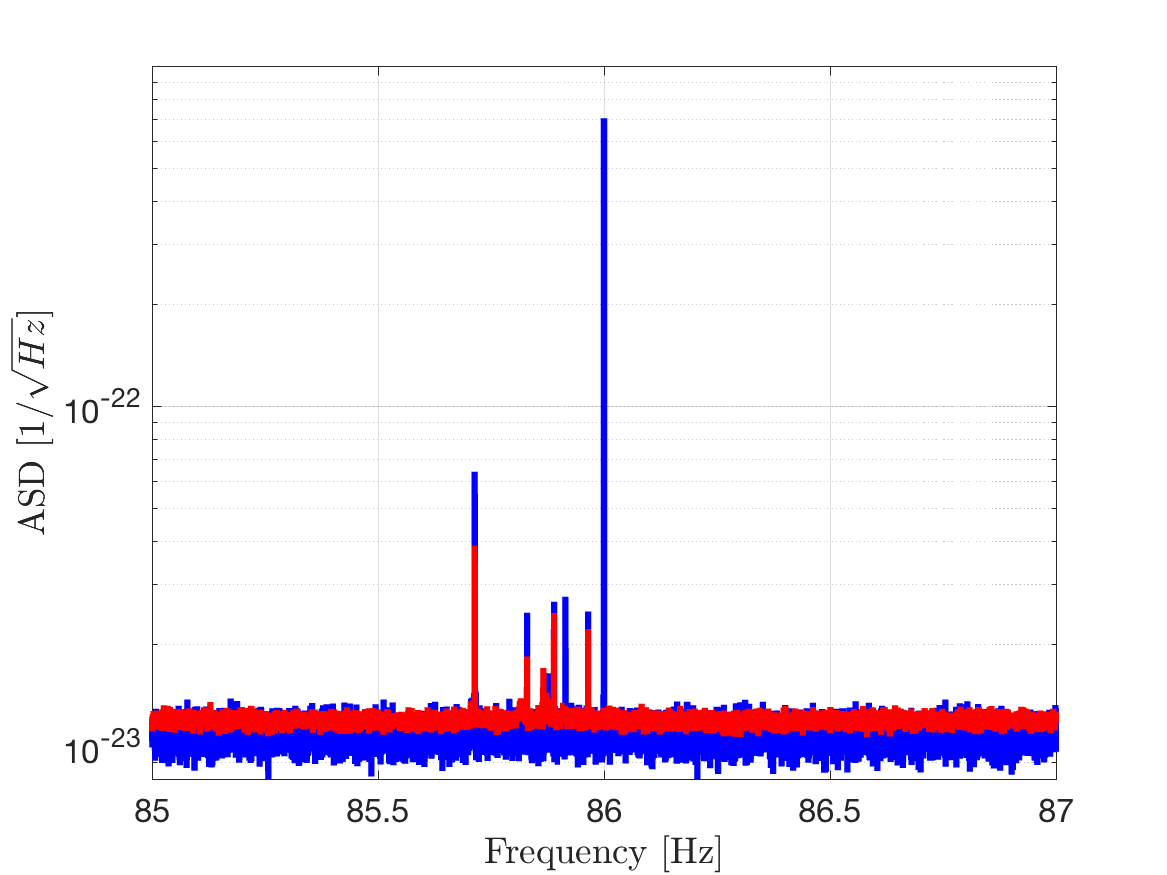}
\caption{\label{fig:86Hzline}Comparison of noise-weighted averaged ASD using data H1 from 8 June 2017 to 15 June 2017 (blue trace, PCAL high-frequency injection on) with noise-weighted averaged ASD from 14 July 2017 to 17 August 2017 (red trace, PCAL high-frequency injection off). The 86~Hz line has disappeared in the second period.}
\end{figure}

A line at 86~Hz was discovered the 15th of June 2017 in the Hanford GW strain data \cite{PCALline1}. After investigating this with a coherence tool, we saw that this line was also present in some photon calibration (Pcal) channels. The Pcal system applies calibrated forces to the end mirrors and is used for interferometer output calibration \cite{PCAL}. The line had appeared for the first time exactly at the same time as the frequency of a Pcal high frequency injection at 5950~Hz was changed. Turning off this injection made the line in the GW channel disappear, as can be seen in figure~\ref{fig:86Hzline}.

While the coupling mechanism remains unclear, a working hypothesis is that the data acquisition system down-converts the high frequency injection to low frequency lines. A phenomenological equation to predict the frequency of the lines was derived: $f_{line}=2^{16}-f_{inj}*n$, where $f_{inj}$ is the frequency of the injection and $n$ is the nth harmonic (the observed line was the 11th harmonic). This equation was tested changing the frequency of the injections, and it predicted correctly the frequency of the lines. After discovering this, a similar line, also produced by down-conversion, was observed in the Livingston GW strain channel at 119.9~Hz \cite{PCALline2}. Down-converted lines due to photon calibrator actuation do not appear appreciably in the GW spectrum above $\sim$150~Hz because the force-to-length transfer function decays as $f^{-2}$.

\section{\label{sec:knownlines}Producing a list of known lines and combs}

In this section we briefly describe how lists of known instrumental lines are generated for each observing run. Different approaches are followed by the CW group and the SGWB group, because the stochastic searches are only affected by lines that are coherent between both detectors i.e. have the same frequency, whereas CW searches are also affected by lines present in only one detector.

The Appendix includes tables summarizing lines and combs that were found in the O1 and O2 data sets, lines deemed safe to veto {\it a priori} in searches.

\subsection{List of known lines and combs for CW searches}
Searches for CWs, such as recent all-sky searches for unknown isolated sources~\cite{CWO1AllSky}, typically use a list of known lines and combs to veto frequency bands prior to running the searches or, afterward, for vetoing outliers. We summarize here the procedure used to generate these lists.

We begin by generating Tukey-windowed short Fourier transforms (SFTs), using the standard FFT code that is part of the LALSuite library~\cite{LALSUITE}. We generate 7200-s-long SFTs for each detector covering all of the observing run time, because those are the lengths of the longest SFTs used in O1 semi-coherent searches based on summing SFT powers. Then, we compute the inverse-noise-weighted average of the SFTs as described in section~\ref{sec:fscan}. 

The lines are found by visual inspection of the spectrum. Features that appear to be above the neighboring noise level are noted for further inspection. Since, in principle, a narrowband astrophysical source might appear in the spectrum, this initial list is not regarded as safe for line vetoing or cleaning.  

After this first pass, we try to correlate the found lines with other channels using the coherence and NoEMi tools, we check which lines belong to previous known combs using the FineTooth tool (and try to find new combs in the data), and we find the lines produced by known sources like the power mains, the calibration lines, and the mechanical resonances of the different suspensions. This allows us to produce a first list with lines that are safe to veto. This list is updated as more investigations are carried out, including detector studies and targeted follow-up of outliers from searches.

Some lines occupy a single frequency bin, while others occupy several bins. For the latter case, we define the width of the line by noting the interval for which there is a statistically significant excess above the background level estimated from neighboring bins. Non-stationary ``wandering lines'' can be narrow in a particular short time epoch, but vary in frequency over an observation run, leading to a substantial widths in a run-averaged spectrum. Visual inspection is used in these studies, rather than fully automated methods, because of the enormous range in line widths encountered, combined with overlapping line structures that challenge automated tools. 

\subsection{List of known lines and combs for SGWB searches}
\label{sec:make_SGWB_notchlist}

Searches for an SGWB typically notch out lines of excess power due to well-known sources immediately. These include violin modes, calibration lines, and any simulated pulsars added by hardware injections~\cite{o1_hwinj}. After this, a multi-step process is used to remove other frequency bins from our analysis.

We begin by taking lines associated with loud coherence between the Hanford and Livingston detectors. In principle, this might include genuine GWs. However, we then cross-check these lines against our strain-auxiliary channel coherences. If we find excess coherence at the same frequency in strain-strain coherence as a strain-auxiliary channel coherence, then we remove this frequency from all cross-correlation based analyses. Often these lines are associated with electronics, and so we see coherence with magnetometers or mains monitors. This is the case for the 8~Hz comb shown Figure~\ref{fig:coherence}.

If we do see coherence in our strain-strain measurement but not in any of our strain-auxiliary channel coherence measurements, then we might employ different strategies depending on the search being performed. For broadband searches that integrate over frequency and the signal model is a power-law in frequency, we might remove these lines. They are not consistent with the proposed signal model. However, for directed, narrow-band searches that look for signals in each individual frequency bin, we might still consider these frequencies in our analysis. Given that these cross-correlation-based searches are for a known direction and the Doppler shift for a source from that direction due to the motion of the Earth is known, high frequency excess coherence in one single bin is often inconsistent with the signal model for a persistent source in that direction. Therefore, we often remove frequency bins in cases where there is a single very loud frequency bin at a resolution much smaller the expected Doppler broadening of a signal from the direction in which we are looking. This is the case for many of the single frequencies marked "Unknown" in Table~\ref{tab:SGWB_O1_notchlist}.

Finally, we employ the comb-finder described in section~\ref{sssec:coherence_comb_finder} to find any obvious combs in our data that might not be evident from excess power statistics, but might be prevalent enough to cause excess broadband signal in our final detection statistic. If we find convincing evidence for a comb then we remove all potential comb teeth from the analysis. The comb finder was used to notch a 1 Hz comb in O1, as shown in Figure~\ref{fig:O1comb}.

\section{\label{sec:conclusions}Conclusions}

We have demonstrated the methods used for identification of narrow spectral artifacts caused by non-astrophysical disturbances. These efforts benefit searches for persistent gravitational wave signals by identifying those frequency bands affected by such disturbances. Some artifacts are caused by sources that can be mitigated. Several examples of such mitigation efforts have been presented. While some of the most pervasive combs have been reduced or mitigated, the causes of other artifacts remain unknown (see figure~\ref{fig:spectra} and tables in the appendix).

Between the second and third Advanced LIGO observing runs, a series of upgrades and other improvements are under way, in order to bring the detectors closer to their design sensitivities. As detector noise is reduced, other, previously unseen lines and combs are likely to become apparent, requiring further identification and mitigation efforts. In addition, as described in this article, detector maintenance activities can inadvertently create new spectral artifacts. Careful monitoring of the data will continue to be required in order to prevent contamination of long epochs of detector data. Mitigating narrow spectral artifacts will also be needed well into the advanced gravitational wave detector era.

\begin{acknowledgments}
The authors gratefully acknowledge the support of the United States
National Science Foundation (NSF) for the construction and operation of the
LIGO Laboratory and Advanced LIGO as well as the Science and Technology Facilities Council (STFC) of the
United Kingdom, the Max-Planck-Society (MPS), and the State of
Niedersachsen/Germany for support of the construction of Advanced LIGO 
and construction and operation of the GEO600 detector. 
Additional support for Advanced LIGO was provided by the Australian Research Council.
The authors gratefully acknowledge the Italian Istituto Nazionale di Fisica Nucleare (INFN),  
the French Centre National de la Recherche Scientifique (CNRS) and
the Foundation for Fundamental Research on Matter supported by the Netherlands Organisation for Scientific Research, 
for the construction and operation of the Virgo detector
and the creation and support  of the EGO consortium. 
The authors also gratefully acknowledge research support from these agencies as well as by 
the Council of Scientific and Industrial Research of India, 
the Department of Science and Technology, India,
the Science \& Engineering Research Board (SERB), India,
the Ministry of Human Resource Development, India,
the Spanish  Agencia Estatal de Investigaci\'on,
the Vicepresid\`encia i Conselleria d'Innovaci\'o, Recerca i Turisme and the Conselleria d'Educaci\'o i Universitat del Govern de les Illes Balears,
the Conselleria d'Educaci\'o, Investigaci\'o, Cultura i Esport de la Generalitat Valenciana,
the National Science Centre of Poland,
the Swiss National Science Foundation (SNSF),
the Russian Foundation for Basic Research, 
the Russian Science Foundation,
the European Commission,
the European Regional Development Funds (ERDF),
the Royal Society, 
the Scottish Funding Council, 
the Scottish Universities Physics Alliance, 
the Hungarian Scientific Research Fund (OTKA),
the Lyon Institute of Origins (LIO),
the Paris \^{I}le-de-France Region, 
the National Research, Development and Innovation Office Hungary (NKFI), 
the National Research Foundation of Korea,
Industry Canada and the Province of Ontario through the Ministry of Economic Development and Innovation, 
the Natural Science and Engineering Research Council Canada,
the Canadian Institute for Advanced Research,
the Brazilian Ministry of Science, Technology, Innovations, and Communications,
the International Center for Theoretical Physics South American Institute for Fundamental Research (ICTP-SAIFR), 
the Research Grants Council of Hong Kong,
the National Natural Science Foundation of China (NSFC),
the Leverhulme Trust, 
the Research Corporation, 
the Ministry of Science and Technology (MOST), Taiwan
and
the Kavli Foundation.
The authors gratefully acknowledge the support of the NSF, STFC, MPS, INFN, CNRS and the
State of Niedersachsen/Germany for provision of computational resources.
The authors also gratefully acknowledge the support of the LIGO Scientific Collaboration Fellows program.
This article has LIGO document number P1700440.
\end{acknowledgments}

\appendix*
\section{Known lines and combs for O1 and O2}

We present a table of lines and a table of combs for the O1 and O2 data runs, with a description of the source of the noise in each case for which it is known\footnote{Updated and more detailed lists can be found at \url{https://losc-dev.ligo.org/o1speclines/}}. Table \ref{tab:combs_O1O2} shows a list of O1 and O2 combs that have been identified at the time of this writing, while table \ref{tab:lines_O1O2} shows a list of O1 and O2 single lines which do not belong to any known comb.

\begin{table*}[htbp]
\begin{center}
\scriptsize
\begin{tabular}{cccccc}
\hline
Spacing & Offset & Range of & Description & Detector & Run\\
(Hz) & (Hz) & visible harmonics & & & \\
\hline \hline \\
0.0470* & 972.1417 & 0-1 & Unknown & H1 & O2 \\
0.088425* & 76.3235 & 0-14 & Unknown & H1 & O1 \\
0.08844* & 153.4428 & 0-9 & Unknown & H1 & O2 \\
0.2000	 & 0.0000 & 106-191 & Unknown & L1 & O2 \\
0.6000* & 0.5690 & 742-745 & Unknown & L1 & O2 \\
0.9865* & 18.7433 & 0-37 & Unknown & L1 & O1 \\
0.9878881 & 0.0000 & 21-64 & Unknown & H1 & O2 \\
0.987925 & 0.0000 & 25-52 & Unknown & L1 & O2 \\
0.98793 & 21.7344 & 0-27 & Unknown & L1 & O1 \\
0.99678913 & 0.0000 & 23-695 & Unknown & L1 & O2 \\
0.9967943 & 0.0000 & 21-685 & Unknown & L1 & O2 \\
0.99816 & 30.9430 & 0-30 & Unknown & H1 & O1 \\
0.9981625 & 64.8804 & 0-8 & Unknown & H1 & O1 \\
0.9991573 & 0.0000 & 26-89 & Unknown & H1 & O1 \\
0.999970 & 18.2502 & 0-35 & Unknown & L1 & O1 \\
0.999975 & 76.75 & 0-36 & Unknown & L1 & O1 \\
0.999979* & 31.7512 & 0-24 & Unknown & L1 & O1 \\
0.9999862 & 0.2503172 & 20-52 & Unknown & H1 & O2 \\
0.999989 & 20.5000 & 0-69 & Unknown & L1 & O1 \\
0.99999 & 19.2500 & 0-33 & Unknown & H1 & O1 \\
1.0000 & 0.0000 & 20-140/20-125 & Unknown & L1/H1 & O2 \\
1.0000* & 15.7487 & 0-13 & Unknown & L1 & O1 \\
1.0000 & 16.0000 & 0-94 & Unknown & H1 & O1 \\
1.0000 & 8.5000 & 0-136 & Blinking LEDs in timing system & H1 & O1 \\
1.0000 & 0.1000 & 1238-1416 & Unknown & L1 & O2 \\
1.0000 & 0.5000 & 20-77/20-136 & Unknown & L1/H1 & O2 \\
1.0000 & 0.9987 & 23-114 & Unknown & H1 & O2 \\
1.0000 & 0.9994 & 20-43 & Unknown & H1 & O2 \\
1.4311 & 40.0737 & 0-5 & Unknown & L1 & O1 \\
1.7000 & 0.3500 & 25-31 & Unknown & L1 & O2 \\
1.9464* & 9.73203 & 0-27 & Unknown & L1 & O1 \\
2.040388 & 0.0000 & 9-34 & Unknown & H1 & O1 \\
2.074121875 & 0.0000 & 9-32 & Unknown & H1 & O1 \\
2.074231250 & 0.0000 & 9-32 & Unknown & H1 & O1 \\
2.109236 & 0.0000 & 14-30 & Unknown & H1 & O2 \\
2.202136 & 0.0000 & 11-22 & Unknown & L1 & O1 \\
2.20458 & 0.0000 & 10-21 & Unknown & L1 & O1 \\
3.89284 & 37.0226 & 0-5 & Unknown & L1 & O1 \\
4.0000 & 27.7633 & 0-4 & Unknown & L1 & O1 \\
8.0000 & 0.0000 & 1-250 & OMC length dither & H1 & O1\\
8.0000 & 0.0000 & 3-16 & Unknown & H1 & O2 \\
11.1111 & 0.0000 & 1-6 & Vacuum sensors & H1 & O2 \\
11.394784 & 0.0000 & 2-8 & Unknown & H1 & O2 \\
11.395279 & 0.0000 & 2-8 & Unknown & H1 & O2 \\
11.92117 & 19.8422 & 0-6 & Unknown & L1 & O1 \\
11.985395 & 0.0000 & 1-22 & Unknown & L1 & O1 \\
19.07328 & 9.53672 & 1-7 & Unknown & H1 & O2 \\
20.83272 & 0.0000 & 1-46 & Unknown & H1 & O1 \\
31.4127 & 0.0000 & 1-2 & Unknown & H1 & O1 \\
31.4149 & 0.0000 & 1-2 & Unknown & H1 & O1 \\
56.840557 & 0.0000 & 1-7 & Unknown & H1 & O1 \\
60.0000 & 0.0000 & 1-9 & Power mains & H1/L1 & O1/O2 \\
66.665 & 0.0000 & 1-2 & Unknown & L1 & O1 \\
76.32344 & 0.0000 & 1-8 & Unknown & H1 & O1 \\
99.9987 & 0.0000 & 1-7 & Unknown & H1 & O2 \\
99.99877 & 0.0000 & 1-12 & Unknown & H1 & O1/O2 \\
99.99925625 & 0.0000 & 4-20 & Unknown & L1 & O1 \\
99.99928 & 0.0000 & 1-20 & Unknown & L1 & O1 \\
\hline
\end{tabular}
\caption{\label{tab:combs_O1O2}All identified combs at the time of this writing during O1 and O2 that appeared in the run-averaged spectra (spacings marked with a * produced more than one comb with different offsets and showing at different harmonics). The frequencies of the teeth of a comb are given by: $f_n = f_{o} + n*\delta f$, where $f_{o}$ is given by the second column, $\delta f$ is given by the first column and $n$ is given by the third column. Most of the identified combs are from unknown origin and have not been eliminated at the time of this writing.}
\end{center}
\end{table*}

\begin{table*}[htbp]
\begin{center}
\scriptsize
\begin{tabular}{cccc}
\hline
Freq. (Hz) & Description & Detector & Run \\
\hline \hline \\
28.6100   & Coherent with safe PEM channel(s) & H1 & O1/O2    \\
29.8019   & Coherent with safe PEM channel(s) & H1 & O1/O2    \\
35.7048   & Coherent with safe PEM channel(s) & H1 & O1/O2    \\
35.7065   & Coherent with safe PEM channel(s) & H1 & O1/O2    \\
35.7624   & Coherent with safe PEM channel(s) & H1 & O1/O2    \\
35.7628   & Coherent with safe PEM channel(s) & H1 & O1/O2    \\
35.9000   & Calibration & H1 & O1/O2 \\
36.7000   & Calibration & H1 & O1/O2 \\
37.3000   & Calibration & H1 & O1/O2 \\
44.7029   & Coherent with safe PEM channel(s) & H1 & O1    \\
59.5110   & Coherent with safe PEM channel(s) & H1 & O1    \\
59.5229   & Coherent with safe PEM channel(s) & H1 & O1    \\
74.5049   & Coherent with safe PEM channel(s) & H1 & O1    \\
83.3155   & Coherent with safe PEM channel(s) & H1 & O1    \\
89.4060   & Coherent with safe PEM channel(s) & H1 & O1    \\
99.9790   & Coherent with safe PEM channel(s) & H1 & O1/O2    \\
104.3068  & Coherent with safe PEM channel(s) & H1 & O1    \\
299.60    & Beam-splitter violin mode & H1 & O1/O2 \\
302.22    & Beam-splitter violin mode & H1 & O1/O2 \\
303.31    & Beam-splitter violin mode & H1 & O1/O2 \\
331.9000  & Calibration & H1 & O1/O2 \\
495-513 & Test mass violin mode region & H1 & O1/O2 \\
599.14    & Beam-splitter violin mode & H1 & O1/O2 \\
599.42    & Beam-splitter violin mode & H1 & O1/O2 \\
604.49    & Beam-splitter violin mode & H1 & O1/O2 \\
606.67    & Beam-splitter violin mode & H1 & O1/O2 \\
898.78    & Beam-splitter violin mode & H1 & O1/O2 \\
899.24    & Beam-splitter violin mode & H1 & O1/O2 \\
906.83    & Beam-splitter violin mode & H1 & O1/O2 \\
910.10    & Beam-splitter violin mode & H1 & O1/O2 \\
986-1014 & Test mass violin mode region & H1 & O1/O2 \\
1083.7000 & Calibration & H1 & O1/O2 \\
1456-1488 & Test mass violin mode region & H1 & O1/O2 \\
1922-1959 & Test mass violin mode region & H1 & O1/O2 \\
\hline
\end{tabular}
\qquad
\begin{tabular}{cccc}
\hline
Freq. (Hz) & Description & Detector & Run \\
\hline \hline \\
22.7000   & Calibration & L1 & O2    \\
23.3000   & Calibration & L1 & O2    \\
23.9000   & Calibration & L1 & O2    \\
31.5118   & Coherent with safe PEM channel(s) & L1 & O1    \\
33.7000   & Calibration & L1 & O1    \\
34.7000   & Calibration & L1 & O1    \\
35.3000   & Calibration & L1 & O1    \\
35.7064   & Coherent with safe PEM channel(s) & L1 & O1    \\
35.7632   & Coherent with safe PEM channel(s) & L1 & O1    \\
39.7632   & Coherent with safe PEM channel(s) & L1 & O1    \\
99.9775   & Coherent with safe PEM channel(s) & L1 & O1    \\
100.0000  & Coherent with safe PEM channel(s) & L1 & O1    \\
100.0020  & Coherent with safe PEM channel(s) & L1 & O1    \\
306.20     & Beam-splitter violin mode & L1 & O1/O2 \\
307.34    & Beam-splitter violin mode & L1 & O1/O2 \\
307.50     & Beam-splitter violin mode & L1 & O1/O2 \\
315.10     & Beam-splitter violin mode & L1 & O1/O2 \\
331.3000  & Calibration & L1 & O1/O2 \\
333.33    & Beam-splitter violin mode & L1 & O1/O2 \\
497-520 & Test mass violin mode region & L1 & O1/O2 \\
615.03     & Beam-splitter violin mode & L1 & O1/O2 \\
629.89    & Beam-splitter violin mode & L1 & O1/O2 \\
630.17     & Beam-splitter violin mode & L1 & O1/O2 \\
630.39     & Beam-splitter violin mode & L1 & O1/O2 \\
918.76    & Beam-splitter violin mode & L1 & O1/O2 \\
926.63     & Beam-splitter violin mode & L1 & O1/O2 \\
945.35    & Beam-splitter violin mode & L1 & O1/O2 \\
945.72     & Beam-splitter violin mode & L1 & O1/O2 \\
991-1030 & Test mass violin mode region & L1 & O1/O2 \\
1083.1000 & Calibration & L1 & O1/O2 \\
1225.20     & Beam-splitter violin mode & L1 & O1/O2 \\
1457-1512 & Test mass violin mode region & L1 & O1/O2 \\
1922-1990 & Test mass violin mode region & L1 & O1/O2 \\
\hline
\end{tabular}
\caption{\label{tab:lines_O1O2}Some known lines from O1 and O2 which do not belong to any found comb. Many more lines are found in the run-averaged spectra, but only lines from known origin or also found in other channels are reported as being safe to veto by the astrophysical searches.}
\end{center}
\end{table*}

\begin{table*}[htbp]
\begin{center}
\scriptsize
\begin{tabular}{c c c}
\hline
Line or Comb & Frequency (Hz) & Description \\
\hline \hline \\
Comb & Offset 0.5 Hz, Spacing 1 Hz & 1 Hz comb \\
Comb & Offset 0 Hz, Spacing 16 Hz & 16 Hz comb \\
Comb & Offset 60 Hz, Spacing 1 Hz & Power mains \\
Line & 20.22 & Unknown\\ 
Line & 20.40 & Unknown\\ 
Line & 23.36 & Unknown\\ 
Line & 24.25 & Unknown \\
Line & 25.00 & Unknown \\
Line & 26.17 & Unknown \\ 
Line & 30.00 & Unknown \\ 
Line & 47.69 & Unknown\\ 
Line & 100.00 & Unknown \\ 
Line & 453.32 & Unknown \\ 
Line & 1352.90 & Unknown\\  
Line & 34.7 & Calibration (L1) \\
Line & 35.3 & Calibration (L1) \\
Line & 36.7 & Calibration (H1) \\
Line & 37.3 & Calibration (H1) \\
Line & 331.3 & Calibration (L1) \\
Line & 331.9 & Calibration (H1) \\
Line & 1083.1 & Calibration (L1) \\
Line & 1083.7 & Calibration (H1) \\
Line & 3001.1 & Calibration (L1) \\
Line & 3001.3 & Calibration (H1) \\
Line & 480-520 & Violin mode first harmonic region \\
Line & 960-1040 & Violin mode second harmonic region \\
Line & 1455-1540 & Violin mode third harmonic region \\
Line & 1200-1300 & Wandering line at Hanford \\
Line & 12.43 & Pulsar injection \\
Line & 26.34 & Pulsar injection \\
Line & 31.42-31.43 & Pulsar injection \\
Line & 38.43-38.51 & Pulsar injection \\
Line & 52.80-52.81 & Pulsar injection \\
Line & 108.85-108.87 & Pulsar injection \\
Line & 146.11-146.21 & Pulsar injection \\
Line & 190.95-191.09 & Pulsar injection \\
Line & 575.11-575.22 & Pulsar injection \\
Line & 265.55-265.60 & Pulsar injection \\
Line & 763.77-763.92 & Pulsar injection \\
Line & 848.88-849.06 & Pulsar injection \\
Line & 1220.43-1220.68 & Pulsar injection \\
Line & 1393.23-1393.79 & Pulsar injection \\
\hline
\end{tabular}
\caption{\label{tab:SGWB_O1_notchlist}Notch list used in SGWB searches for O1. This table lists the frequencies which were not analyzed in SGWB searches in O1 because they were determined to have strong instrumental contamination, following the procedure in section~\ref{sec:make_SGWB_notchlist}. A 0.1~Hz region around each of the harmonics of the 60~Hz lines coming from the power mains was removed. Frequencies where an injection was performed were removed: calibration lines at each site as well as frequencies with hardware injection simulating pulsars. For the pulsar injections, we account for the Doppler shift and the spin-down of the pulsar  over the course of the run. We remove a broad band around the harmonics of the violin modes because of excess noise in these regions. We also remove a wandering line seen at Hanford. Finally, we remove lines seen as coherent between H1 and L1 which have been determined to contaminated with instrumental artifacts. This includes a comb with 1~Hz spacing and 0.5~Hz offset which was identified using the comb finder.}
\end{center}
\end{table*}

\end{document}